\documentclass[manuscript,screen]{acmart}
\AtBeginDocument{%
  \providecommand\BibTeX{{%
    \normalfont B\kern-0.5em{\scshape i\kern-0.25em b}\kern-0.8em\TeX}}}

\setcopyright{acmcopyright}
\copyrightyear{2023}
\acmYear{2023}
\acmDOI{XXXXXXX.XXXXXXX}

\acmConference[C\&C '24]{Make sure to enter the correct
  conference title from your rights confirmation emai}{June 24--26,
  2024}{Chicago, USA}

\acmPrice{15.00}
\acmISBN{978-1-4503-XXXX-X/18/06}

\usepackage{multicol}
\usepackage{multirow}
\usepackage{array}
\usepackage{cleveref}
\usepackage{subcaption}
\usepackage{xcolor}
\usepackage{enumitem}
\usepackage{multirow}
\usepackage[export]{adjustbox}
\usepackage{subcaption}
\definecolor{ForestGreen}{RGB}{34,139,34}
\definecolor{ao}{rgb}{0.0, 0.0, 1.0}
\usepackage[export]{adjustbox}

\usepackage{makecell}
\usepackage{xspace}

\begin{document}

\title{Creativity Support in the Age of Large Language Models: An Empirical Study Involving Emerging Writers}

\author{Tuhin Chakrabarty$^*$}%\thanks{*Equal contribution.}
\email{tuhin.chakr@cs.columbia.edu}
\affiliation{%
  \institution{Columbia University}
  \country{USA}
}

\author{Vishakh Padmakumar$^*$}
\affiliation{%
  \institution{New York University}
  \country{USA}
}

\author{Faeze Brahman}
\affiliation{%
  \institution{Allen Institute for Artificial Intelligence}
  \country{USA}
}

\author{Smaranda Muresan}
\email{smara@cs.columbia.edu}
\affiliation{%
  \institution{Columbia University}
  \country{USA}
}

\renewcommand{\shortauthors}{Chakrabarty, et al.}
\renewcommand{\shortauthors}{Chakrabarty, et al.}

\begin{abstract}

The development of large language models (LLMs) capable of following instructions and engaging in conversational interactions sparked increased interest in their utilization across various support tools. We investigate the utility of modern LLMs in assisting professional writers via an empirical user study (n=30). The \textit{design of our collaborative writing interface} is grounded in the \textit{cognitive process model of writing}~\cite{flower1981cognitive} that views writing as a goal-oriented thinking process encompassing non-linear cognitive activities: planning, translating and reviewing. Participants are asked to submit a post-completion survey to provide detailed feedback on the potential and pitfalls of LLMs as writing collaborators. Upon analyzing the writer-LLM interactions, we find that while writers seek LLM's help across all three types of cognitive activities, they find LLMs more helpful in translation and reviewing. Our findings from analyzing both the interactions and the survey responses highlight future research directions in creative writing assistance using LLMs.
\end{abstract}

\begin{CCSXML}
<ccs2012>
   <concept>
       <concept_id>10003120.10003121.10011748</concept_id>
       <concept_desc>Human-centered computing~Empirical studies in HCI</concept_desc>
       <concept_significance>500</concept_significance>
       </concept>
   <concept>
       <concept_id>10003120.10003130.10011762</concept_id>
       <concept_desc>Human-centered computing~Empirical studies in collaborative and social computing</concept_desc>
       <concept_significance>500</concept_significance>
       </concept>
   <concept>
       <concept_id>10010147.10010178.10010179.10010182</concept_id>
       <concept_desc>Computing methodologies~Natural language generation</concept_desc>
       <concept_significance>300</concept_significance>
       </concept>
 </ccs2012>
\end{CCSXML}

\ccsdesc[500]{Human-centered computing~Empirical studies in HCI}
\ccsdesc[500]{Human-centered computing~Empirical studies in collaborative and social computing}
\ccsdesc[300]{Computing methodologies~Natural language generation}

% Author Keywords
\keywords{Human-AI collaboration, Co-Creativity, Computational Creativity, Large Language Models, StoryTelling, Natural Language Generation, Evaluation, Creativity}% Print the classficiation codes

\newcommand\todo[1]{\textcolor{red}{[TODO: #1]}}
\newcommand\vp[1]{\textcolor{blue}{[VP: #1]}}
\newcommand\fb[1]{\textcolor{brown}{[FB: #1]}}
\newcommand\tc[1]{\textcolor{orange}{[TC: #1]}}
\newcommand\sm[1]{\textcolor{purple}{[SM: #1]}}
% uncomment these to remove comment
%\renewcommand\tc[1]{}
%\renewcommand\fb[1]{}
%\renewcommand\vp[1]{}

\newcommand{\PreserveBackslash}[1]{\let\temp=\\#1\let\\=\temp}
\newcolumntype{C}[1]{>{\PreserveBackslash\centering}p{#1}}
\newcolumntype{R}[1]{>{\PreserveBackslash\raggedleft}p{#1}}
\newcolumntype{L}[1]{>{\PreserveBackslash\raggedright}p{#1}}

\definecolor{red}{RGB}{255, 0, 0}
\definecolor{blue}{RGB}{135, 206, 250}
\definecolor{green}{RGB}{205, 255, 204}

\newcommand{\eg}{e.g.,\xspace}
\newcommand{\ie}{i.e.\xspace}
\begin{teaserfigure}
  \small
  \centering
  \includegraphics[width=0.85\textwidth]{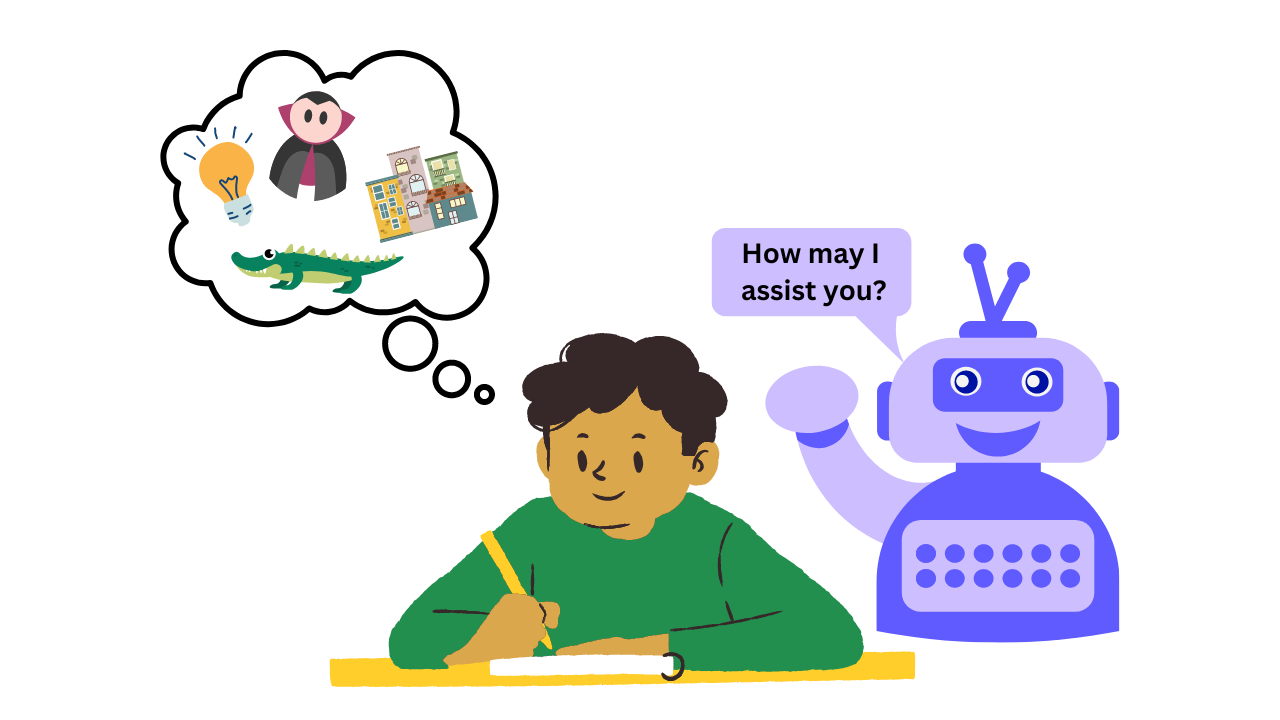}
  \caption{A writer seeks feedback from an AI}
  \label{fig:teaser}
\end{teaserfigure}

\received{20 February 2007}
\received[revised]{12 March 2009}
\received[accepted]{5 June 2009}

\maketitle

%SM I created a new intro file so I can edit, but also keep the old intro if anyone wants to keep it. 
\section{Introduction}
\label{sec:intro}
Writers have often found themselves in the moral quandary of not wanting to use AI but at the same time experimenting with it to understand its potential and limitations. In her essay, Pulitzer fiction finalist Vauhini Vara writes 

\begin{quote} 
    \centering 
    ``We both make our livings as writers and technological capitalism has been exerting a slow suffocation on our craft. A machine capable of doing what we do, at a fraction of the cost, feels like a threat. Yet I found myself irresistibly attracted to GPT-3—to the way it offered, without judgment, to deliver words to a writer who has found herself at a loss for them" \cite{Ghosts} 
\end{quote}

Recent work has demonstrated the promise of large language models (LLMs) in assisting users with a variety of creative writing tasks
from short fiction \cite{yang2022re3,yuan2022wordcraft} to screenplays \cite{mirowski2023cowriting}. Aligning LLMs to generate text more in line with human preferences \cite{ouyang2022training} has enabled them to respond directly to user instructions, transforming them into viable tools for assisting users who do not have a computational background. This has increased the commercial appeal of deploying LLMs as %software tools has led to 
products that offer writing assistance such as \textit{Sudowrite}\footnote{\url{https://www.sudowrite.com/}} that can continue a story where the user left off, describe a scene, rewrite according to some guidelines or even provide feedback. However, the true utility of these models in assisting \textit{professional writers} %is still debatable and 
remains an open question \cite{mirowski2023cowriting, ippolito2022creative}.\footnote{\citet{lee2022coauthor} investigate the capabilities of LLMs as assistive tools in creative and argumentative writing tasks. However, their analysis is grounded in the needs of novice writers recruited from crowd-working platforms. The needs of professional writers may differ and hence warrant further investigation.} 

Our goal is to design a human-AI collaboration framework that can help us better understand the needs of professional writers during the writing process and the potential pitfalls of contemporary LLMs as collaborative creative writing assistants.  \citet{gero-etal-2022-design} argue for a design space for writing support tools that are based on the \textit{cognitive process theory of writing}~\cite{flower1981cognitive} that considers writing as a goal-oriented thinking process encompassing three distinct and non-linear cognitive activities --- \textit{Planning}, \textit{Translation} and \textit{Reviewing}. Planning includes setting goals, brainstorming ideas, and organizing the writers' thoughts. Translation is the process of verbalizing ideas and thoughts. Reviewing includes evaluating and revising what has been written. Through a systematic literature review of recent writing support tools, % using this design space, 
\citet{gero-etal-2022-design} find that none of the prior work %studied tools 
facilitate all three activities, with planning and reviewing being under-studied areas.%  %Inspired by this 

To fill this gap in the literature, we propose a human-AI collaboration framework for writing support that is designed based on the \textit{cognitive process theory of writing} ~\cite{flower1981cognitive} that 
allows a professional writer to seek help from a contemporary LLM (GPT-3.5) \textit{during all three cognitive activities --- planning, translating, and reviewing ---, in a non-linear fashion} (Section \ref{sec:interface}).\footnote{We note that \citet{10.1145/3581641.3584060} use the cognitive model from \citet{flower1981cognitive} for qualitative analysis of their next-phrase suggestions when writing movie reviews. Our work extends this line of work to contemporary instruction-tuned LLMs \cite{wei2021finetuned} that more naturally fit the three phases of the writing process, allowing us to incorporate the theory into the design of the interface itself.}
%. Unlike prior work \cite{mirowski2023cowriting,ippolito2022creative}, and  
Using our proposed %human-AI collaboration 
framework we collect a dataset of $30$ stories, spanning between $1000$ and $4000$ words in length across a wide range of genres, written by $17$ emerging writers with the help of GPT-3.5. In addition, participants were asked to submit a post-completion survey to provide feedback on the potential and pitfalls of LLMs as tools for writing assistance (Section \ref{sec:participant_recruitment}). Along with the stories, we also record the interactions between the writers and GPT-3.5, which consist of paired instances of natural language instructions provided by the writers and the corresponding model-generated responses. This enables us to both identify patterns in instruction usage (for what activities do professional writers seek help from the LLM) and to estimate the model's contribution to the stories (in which of these activities is the LLM most helpful) (\Cref{sec:interaction}). 
We observe that while the writers use the LLM for all stages of creative writing, % --- planning, translation and reviewing, 
they find the model most helpful for \emph{translation-based} subtasks such as targeted rewriting of paragraphs in the text, or \emph{review-based} subtasks such as obtaining feedback on their draft. We also find that writers often switch between these phases non-linearly, validating our design choices with the interface (\Cref{sec:interaction_patterns}).
Finally, we qualitatively analyze the post-completion survey feedback provided by the writers to identify the strengths and weaknesses of the model (\Cref{sec:writer_feedback}). Current models are limiting for professional writers in several ways including their repetitive nature, over-reliance on clichés and tropes, lack of nuance, subtext, or symbolism as well as overly moralistic and predictable endings. While some of these weaknesses could be resolved with better prompting, our writers also highlight broader concerns about the model's inability to generate text related to darker topics, as well as its difficulty in understanding the writer's intent. These findings highlight future directions of research in the training of these models to align with these user needs. On a positive note, writers find the model to be the most helpful as a rewriting tool or feedback provider rather than an original idea generator. 

%\citet{10.1145/3581641.3584060} used Flower and Hayes \cite{flower1981cognitive} cognitive model for qualitative analysis of their LM-assisted writing system. Their system was built to assist amateur writers in writing movie reviews. Unlike them, we directly incorporate the Flower and Hayes model in building our support tool. It should also be noted that our task is fundamentally more challenging than a movie review writing task. Similarly \cite{lee2022coauthor} has shown the capabilities of LLM in creative and argumentative writing tasks. However, due to the advanced qualifications of our participants in creative writing, their expectations and needs are fundamentally different from amateur writers on how they use AI in the writing process. Additionally, there has been a giant leap in the performance of LLM's in terms of how much they can support user instructions \cite{wei2021finetuned,wang2022super}. We take advantage of this instead of utilizing LLM's just as autoregressive next word prediction system like prior works. 
To summarize our contributions, building on recent studies on collaborative writing \cite{ippolito2022creative,mirowski2023cowriting}, our work is the first to study the role of state-of-the-art LLMs as writing tools for professional writers using a theoretically grounded interface. We also collect and release, to our knowledge, the first dataset of co-written stories and human-AI interactions with professional writers %on \url{https://collab-stories.github.io/} 
to foster further research along this direction.\footnote{Throughout the paper, we use LLM/AI/GPT-3.5 interchangeably which refer to the same underlying model.}
 \section{Related Work}
\label{sec:related}
\subsection{Human-AI collaborative writing}

Research in collaborative writing has examined the effects of different suggestion formats. \citet{kannan2016smart} created the Smart Reply in Gmail using an LSTM model, significantly influencing mobile email responses. %Studies by 
\citet{10.1145/3325480.3326573,10.1145/3290605.3300526} found that synonym and metaphor suggestions inspire writers. \citet{10.1145/3172944.3172983} observed a preference for interfaces allowing greater user control in collaborative writing. \citet{buschek2021impact} explored the balance between the number of suggestions and their usefulness in enhancing writing efficiency.
More recent work investigates using language models %Research on natural language generation for 
for narrative crafting including story assistance for Japanese novelists \cite{10.1145/3411763.3450391}, a completion-generation model for collaborative story writing \cite{akoury2020storium}, and AI in interactive game narratives \cite{10.1145/3402942.3409599,volum2022craft,akoury2023grounded}. Chatbots have been explored for creating fictional characters \cite{10.1145/3450741.3465253}. Most closely related to our work are Wordcraft \cite{yuan2022wordcraft} and Dramatron \cite{mirowski2023cowriting} which utilize LLMs for various writing tasks, including story continuations and screenplays.
Unlike these works, our interface design is based on the cognitive process theory of writing and we use publicly-available LLMs, with a longer context window, which allows us to share the collected interactions and feedback from users to inspire future research in this direction. 

\subsection{Creativity Support Tools}
Computational writing support has evolved from early spellcheckers \cite{10.1145/3411764.3445645} to recent AI-driven story generation tools \cite{10.1145/3491102.3501819}. Techniques inspired by expert authors enable crowdworkers to complete complex writing tasks via reflection and revision \cite{10.1145/2998181.2998196}. Other advancements include methods for identifying writing style \cite{10.1145/3313831.3376730}, cognitive writing theories for creating writing scaffolds \cite{10.1145/3173574.3173596}, and ideation support tools \cite{10.1145/3325480.3326573,10.1145/3290605.3300526}. Recently, large pre-trained models have emerged as creativity aids for writers \cite{yuan2022wordcraft,mirowski2023cowriting} and visual artists \cite{10.1145/3581641.3584078}.

\citet{51803} emphasize the importance of connected design spaces for exploring the interaction between HCI and pre-trained AI models, influencing prototyping. \citet{gero-etal-2022-design} develop a space based on the cognitive process model of writing, affecting interface design choices. \citet{10.1145/3461778.3462050} review literature on creativity support tools in various arts to create a taxonomy encompassing roles, interactions, and technologies. \citet{10.1145/3491102.3501933,10.1145/3290605.3300619} showed how creative practitioners prioritize functionality, workflow integration, and performance, among other concerns, %UI/UX, learning support, cost, and emotional connection 
when adopting new creativity support tools, 
with most discovering tools via personal recommendations. %Recent studies delve into 
\citet{gero2023social} examine the social dynamics of AI in creative tasks, observing the mismatch between writers' goals and computer support. %Research by 
\citet{10.1145/3532106.3533506} finds that writers value emotional connection and self-confidence, fearing AI might undermine their control and uniqueness. A study with 30 participants by \citet{10.1145/3411764.3445219} explores AI's roles in creative collaboration, like fostering common ground and reducing friction.  \citet{10.1145/3581641.3584060} used the cognitive process model of writing from \citet{flower1981cognitive} for qualitative analysis of their LM-assisted writing system for the task of movie review writing. Unlike them, we directly incorporate the theory from \citet{flower1981cognitive} into %the building of our interface. %SM-cc slight edited
designing and building our interface for creativity support.  

Our work bridges several of these themes across creativity support research. We implement the recommendations from prior work into a theoretically inspired user interface and use it to evaluate LLMs as tools with professional writers. %create a design space that fuses the cognitive process model of writing with large-scale pre-trained models. Further, we release data and provide an in-depth analysis that demonstrates how writers use our tool for planning, translation, and reviewing. 
Finally, our findings highlights several key discussion points, including documenting the desires from writers about AI assistance in creative writing and lines of future research for model training.

\subsection{Generation with large language models}
In the last five years, natural language generation has greatly evolved, primarily owing to larger language models \cite{NEURIPS2020_1457c0d6} that % show improved performance due to our understanding of 
utilize our improved understanding of the scaling laws involved \cite{Rae2021ScalingLM, wei2022emergent} and the advent of alignment training with collected human preferences \cite{ouyang2022training}. %These models, though similar to smaller versions, significantly improve with more data and scale \cite{Rae2021ScalingLM, wei2022emergent}. 
This enables them to %ey now 
perform diverse tasks effectively with little extra training, resulting in %. Many new 
writing tools built using them %se sophisticated models 
\cite{10.1145/3532106.3533533,lee2022coauthor}. However, contemporary LLMs still lack a concrete fact-verification mechanism, leading to potential inaccuracies, especially in non-fiction writing \cite{jiang-etal-2020-know}. They can generate harmful content from seemingly harmless prompts \cite{perez-etal-2022-red,gehman-etal-2020-realtoxicityprompts}, risking misrepresentation of minorities. Ethical concerns also arise, including environmental and socio-political issues \cite{10.1145/3442188.3445922}. Prompt engineering is emerging to manage these challenges \cite{10.1145/3560815,bach-etal-2022-promptsource}, but its complexity may be daunting for non-technical users. % Ethical concerns also arise, including environmental and socio-political issues \cite{10.1145/3442188.3445922}. %Finally, there have been recent AI safety and aligning language models with human preferences through Reinforcement Learning through Human Feedback (RLHF) \cite{christiano2017deep}. %However, these techniques might hinder fiction writers from navigating more complex and darker themes.

% \cite{lee2022coauthor}
% \cite{shen2023parachute}
\section{Building a Collaborative Writing Platform}
\label{sec:interface}
\begin{figure*}
    \centering
    \includegraphics[width=\textwidth]{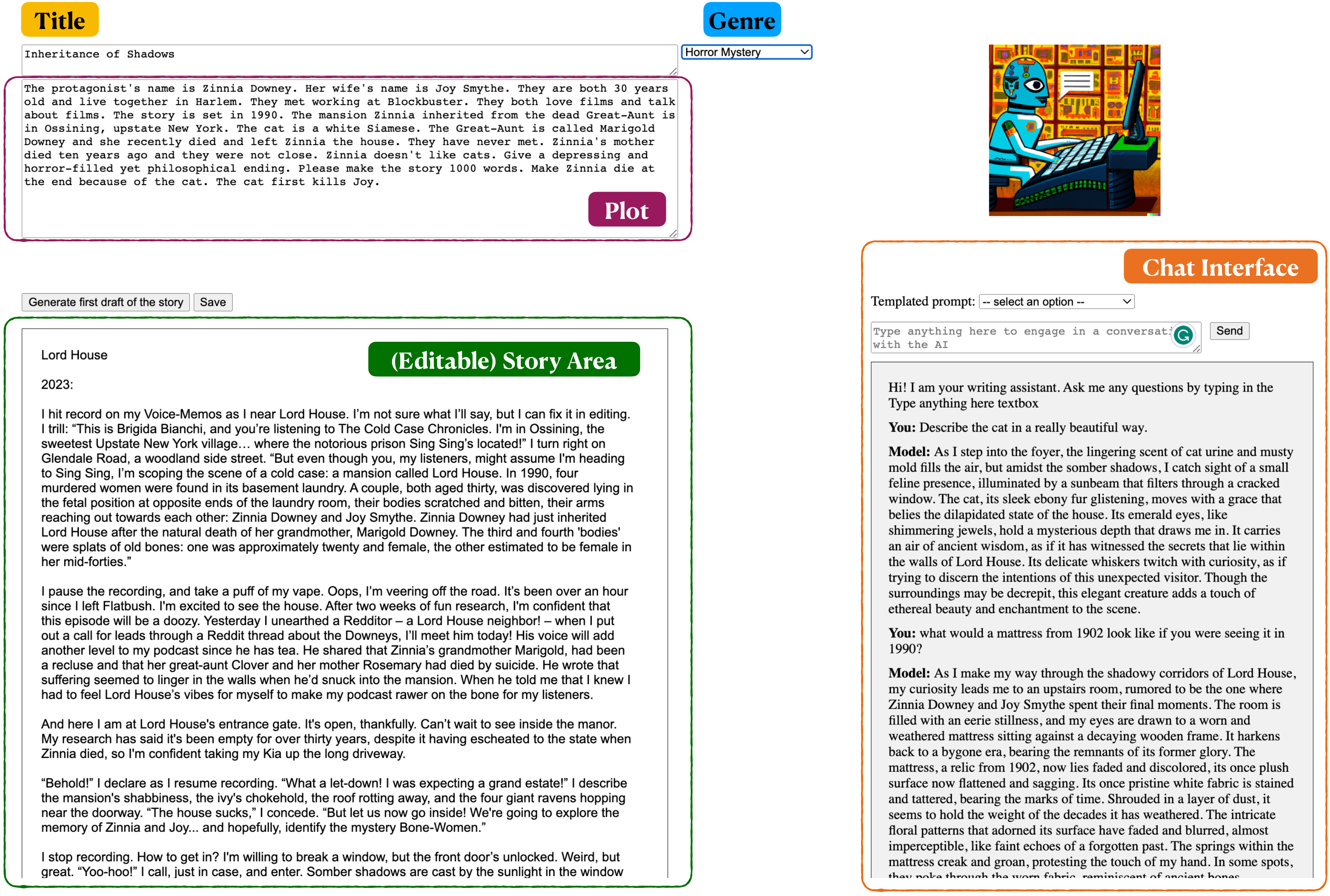}
    \caption{\label{fig:interface}Interaction interface to write a story by collaborating with LLM}
\end{figure*}

Our design for creating a collaborative writing tool is based on prior work that conceptualizes design spaces for creativity support tools \cite{gero-etal-2022-design, 10.1145/3461778.3462050}. %Focusing particularly on writing support tools, 
\citet{gero-etal-2022-design} argue for a design space based on the cognitive process theory of writing \cite{flower1981cognitive} that sees writing as a goal-oriented thinking process that involves three distinct and non-linear cognitive activities: \textit{planning, translating} and \textit{reviewing}. In this section, we detail how our design choices were influenced by this theory allowing the writer to seek help from the LLM in all three non-linear cognitive activities. The overall interaction interface is given in Figure \ref{fig:interface}.
Unlike prior work in co-creative writing, %such as Dramatron \cite{mirowski2023cowriting}, 
which only supports a pre-defined set of domain-specific instructions \cite{mirowski2023cowriting} or provides assistance via auto-complete \cite{lee2022coauthor, 10.1145/3581641.3584060} %catering to script writing primarily involving \textit{Planning}, 
our interface takes advantage of state-of-the-art language models trained on human-written instructions. Hence, they support any freeform user-input %provided natural language instructions 
providing assistance to users in %that are capable of 
all three cognitive activities.

\subsection{System Feature 1: {Writers provide the plot, model writes a draft}}
\label{sec:plot_to_draft}
Our user interface was designed with the intent to allow writers to write short fiction/non-fiction (typically between 1500 and 2000 words) based on a genre and 
a plot of their choice. We started with an initial list of 8 genres \textit{Literary Fiction, Science Fiction, Fantasy, Drama, Thriller, Dystopian, Creative Non-Fiction} and \textit{Horror} and allowed writers to blend or input their own genres. Given the user-provided genre and plot, we first generate an initial story draft using GPT-3.5 with the following prompt, ``\texttt{Generate a 1500-word long story of <genre> genre that follows this plot: <plot>}''. As can be seen in Figure \ref{fig:interface}, the text area on the top left shows the title of the story and the text area below it contains the plot. Once a user hits the \textit{Generate first draft of the story} button, the generated story conditioned on the plot appears on the bigger text area below it. For simplicity, we refer this to as the \textit{story area}. All the text areas are editable giving writers full access to add/edit/delete text. By letting participants provide a plot we allow them to indulge in \textit{Planning}.

Our design decision to add the first draft was mostly motivated by the fact that first drafts are not the place to obsess over the perfection of one's writing style. First drafts are for getting lost in the story and honing it later strictly during the editing phase \cite{lamott1994shitty} %Our decision to add the first draft functionality 
which was further corroborated by some of our participants.

\subsection{System Feature 2: Writers interact with GPT-3.5 through a chat interface and receive suggestions based on the chat history and story draft}
\label{sec:draft_chat}

Along with the capability to generate a first draft based on the plot, we also provide the writers with a chatbot as can be seen on the right side of the interface in Figure \ref{fig:interface}. The chatbot generates text from a state-of-the-art LLM, GPT-3.5, in response to natural language instructions. This acts as a conceptual sandbox, fostering an environment conducive to the use of GPT-3.5 for planning, translation and reviewing. We include the text inside the \textit{story area} as a part of the context in order to enable writers to ask questions pertinent to the story (see Table \ref{chatgpt:start}).

\begin{table}[!ht]
\def\arraystretch{1.25}
\centering
\begin{tabular}{|l|l|}
\hline
Role   & Content                                                                                                                        \\ \hline
System & \begin{tabular}[c]{@{}l@{}}You are a creative assistant helping a user improve this Story Draft: \{\{story\}\}\end{tabular} \\ \hline
\multicolumn{2}{|c|}{\{\{Past interactions between the user and model\}\}} \\ \hline
Writer & Rewrite the first paragraph                                                                                                    \\ \hline
\end{tabular}
\vspace{2ex}
\caption{\label{chatgpt:start}An example input prompt to GPT-3.5 where the System input is always passed to ensure the output response is specific to the story in context. We also provide the history of interactions with the model in case the user wants to simply rephrase the previous request. Finally, we have the most recent request that the writer expects the model to generate a response to.}
%\vspace{2ex}
\end{table}

Owing to the fact that LLMs such as GPT-3.5 are fine-tuned on \texttt{<instruction, output>} pairs, the chat is capable of handling any free-form textual instructions from the writers. We expect participants to use this chat for planning, translation, and reviewing. To further assist writers during this collaboration, we also provided them with some templated prompts (\Cref{templated}). The \textit{Generate
Continuation}, \textit{Rewrite with Imagery} or \textit{Elaborate Selection} functionality were inspired by WordCraft \cite{yuan2022wordcraft}. Finally, during initial experiments, we noticed that the text generated by GPT-3.5 was heavy on exposition and barely contained any dialogue. This led us to introduce the \textit{Insert Dialogue/Monologue} template giving authors the chance to explicitly insert dialogue as needed. For instructions that require localized edits in the draft, as shown in Table \ref{templated}, we allow the writers to select the intended draft snippet using `<' and '>' tags as delimiters. In compliance with \textit{Reviewing},  we also introduce the \textit{Get Feedback} prompt which generates a critique of the draft written inside the \textit{story area}. It should be noted that we did not design templated instructions for \textit{Planning} as these instructions tend to be story-specific.

\begin{table}[!ht]
\def\arraystretch{1.25}
\small
\centering
\begin{tabular}{|L{2cm}|L{12.5cm}|}
\hline
Functionality & Prompt \\ \hline\hline
\begin{tabular}[c]{@{}l@{}}Generate\\ Continuation\end{tabular}      & \begin{tabular}[c]{@{}l@{}}Generate a one paragraph continuation for the given story draft.\end{tabular}                                                                                                                                                       \\ \hline
\begin{tabular}[c]{@{}l@{}}Elaborate\\ Selection\end{tabular}        & \begin{tabular}[c]{@{}l@{}}Elaborate the part of the story draft demarcated by '\textless{}' and '\textgreater{}'. Add more details and return only the edited \\text between those tokens.\end{tabular}                                                       \\ \hline
\begin{tabular}[c]{@{}l@{}}Rewrite with\\ imagery\end{tabular}       & \begin{tabular}[c]{@{}l@{}}Rewrite the part of the story draft demarcated by '\textless{}' and '\textgreater{}' with more rich imagery and return only the\\ edited text between those tokens\end{tabular}                                                    \\ \hline
\begin{tabular}[c]{@{}l@{}}Insert Dialogue\\ / Monologue\end{tabular} & \begin{tabular}[c]{@{}l@{}}Given the story draft insert dialogue or monologue in between some sentences in the part of the story \\ draft demarcated by '\textless{}' and '\textgreater{}' and return only the\\ edited text between those tokens\end{tabular} \\ \hline
Get Feedback                                                         & \begin{tabular}[c]{@{}l@{}}Given the story write some critiques or inconsistencies citing portions. Give your reasons\end{tabular}                                                                                                                             \\ \hline
\end{tabular}
\vspace{2ex}
\caption{\label{templated}Templated prompts provided to writers that can be used as input for any turn during the conversation with the chatbot}
\end{table}

\subsection{System Feature 3: {Large Language Model Choice}}
\label{sec:model_choice}
We select GPT-3.5 over alternative LLMs for both the generation of the initial draft and for generating responses from the chatbot for two reasons. From an economic standpoint, out of OpenAI's public state-of-the-art model offerings\footnote{At the time of our study, GPT-3.5 and GPT-4 ranked in the top-10 models as per \citet{liang2022holistic}.}, the cost per thousand tokens for GPT-3.5 is \$0.002, a rate that is approximately 30 times more cost-efficient %substantially lower 
than the \$0.06 per thousand tokens of GPT-4. %, thereby rendering GPT-3.5 to be approximately 30 times more cost-efficient. 
Secondly, our preliminary experiments showed that GPT-3.5 exhibited a significantly lower computational latency than GPT-4. User feedback on the OpenAI community forum has also noted that GPT-4 incurs a fourfold increase in response time as compared to GPT-3.5.\footnote{\url{https://community.openai.com/t/gpt-4-extremely-slow-compared-to-3-5/106298}} As we aim to help writers obtain model help in real-time, the optimization of response time was deemed a critical requirement.  %Consequently, the aforementioned characteristics of GPT-3.5 prompted us to adopt this particular model in our study.

\subsection{System Feature 4: {Handling the token length limits of GPT-3.5}}
One concern with using GPT-3.5 for model assistance is that we are constrained to use only $4096$ tokens to represent both the context and generated completion.  As a result, we must devise strategic approaches to contain our conversational context within this constraint, while not limiting the utility for the end users. A central feature of our system is the incorporation of the story draft within the system prompt (Table \ref{chatgpt:start}), which enables the writers to formulate instructions pertaining to the story. Hence we could run into this token limit as writers interact with the chatbot over multiple turns. 
We advise users to craft stories that fall within the range of $1500$ to $2000$ words, thereby ensuring that an ample quantity of tokens remains available for %the chatbot to execute 
a constructive dialogue with the model. In the event that users' draft and %sustain a multi-turn conversation to the extent of 
dialogue turns exhaust the $4096$-token limit, we truncate all previous conversational exchanges with the exception of the most recent two turns. The new context, subsequently, consists of the system prompt with the story draft and the two most recent turns from the discourse. This design choice was made by the observed user behavior, where reference to early dialogue exchanges is infrequent. Users predominantly request autonomous instructions or seek guidance contingent on the context established in the immediately preceding conversational turns.

\section{Methodology}
\label{sec:participant_recruitment}

\subsection{Participant Recruitment}
\subsubsection{Recruiting Procedure}
\citet{gero2023social} has argued that the definition of an ‘expert’ or ‘amateur’ creative writer is difficult in a field that has unclear professional delineations. While it is easy to recruit anyone who identifies themselves as a creative writer, our initial experiments revealed a significant variance in the quality of writing. This was partially due to the fact that often content writers without significant experience in writing creative fiction or non-fiction sign up for creative writing tasks on freelancing websites such as \textit{Upwork}, \textit{Prometric} or \textit{User Interviews}. To avoid this, we restricted our recruitment to only those with a formal creative writing education (e.g., a Master of Fine Arts in Creative Writing). Participants were recruited through MFA graduate school distribution lists. The authors of the paper scheduled short Zoom interviews with interested participants explaining the process and asking for their consent to participate in the study. All participants were also notified that their interactions with the AI would be recorded and the final stories written by them would be published. The motivation around this was to incentivize them to write better-quality stories. We additionally used snowball sampling, asking those we found to introduce us to other MFA students. 

\subsubsection{Ensuring Diverse Coverage of Genres}
We hypothesize that the kind of assistance sought out by users when writing different genres of fiction would vary significantly. For example, a dystopian writer may have different expectations about %setting and 
world-building compared to a writer who primarily restricts themselves to literary fiction. A non-fiction writer might be interested more in the attribution of facts generated by LLMs compared to imagery or dialogue. We seek to capture as much of the spectrum of these concerns as possible by recruiting writers who specialize in multiple different genres. In addition,  we recruit participants who are familiar with the use of generative AI and understand, at a surface level, how large language models work to ensure that they were able to fit the model into their existing writing workflow. 
%This allowed us to gain insight into the dynamics of how people think about large language model-based writing assistance. 
We aimed to collect 30 stories covering diverse genres and continued recruiting until we had sufficient participants to reach this target. 

\begin{table*}[!ht]
\centering
\small
\def\arraystretch{1.05}
\begin{tabular}{|c|c|c|c|c|c|c|c|c|c|c|c|c|c|c|c|c|c|c|c|}
\hline
Abbrv & MG & SH & MC & ES & ATF & KK & REB & CK & ATK & NM & LHS & MK & FV & GS & AF & RH & Anon \\ \hline
Count & 2 & 3 & 3 & 3 & 3 & 1 & 1 & 1 & 1 & 2 & 1 & 1 & 2 & 1 & 1 & 1 & 3 \\ \hline
\end{tabular}
\vspace{2ex}
\caption{\label{participants} Participants from 5 MFA Writing programs in the US, were recruited for our collaborative story-writing experiment. The first row represents the acronym for each participant and the second row shows the count of stories written by each participant}
\end{table*}

\subsubsection{Participant Details}
\label{sec:user_recruiting}
Table \ref{participants} shows the $17$ MFA students recruited for our study. Our participants come from $5$ highly-ranked MFA programs in the United States. To further account for the fact that the cost of these programs can be preventative for many people, we attempted to recruit a few people spanning across both fully funded
and self-funded programs. Our recruitment thus resulted in participants, who have been professionally trained in creative writing including some of them soon submitting their first novella for publication, having prior experience teaching short story workshops at the university level, and even nominees of the prestigious Pushcart Prize.\footnote{\url{http://pushcartprize.com/}} Out of the $17$ participants, $13$ of them are MFA Fiction students. ATF is a MFA Non-Fiction student while CK, KK, and REB are pursuing their MFA in Poetry. Due to our recruiting procedure, all our participants wrote, for the most part, in English. The demographics of the participants were: $5$ women, $9$ men, and $3$ non-binary. Our participants were aged between $25$ and $48$.

\subsection{Task Walkthrough}
Once recruited, participants are then expected to interact with GPT-3.5 to co-write a short fiction in a total allotted time of $3$ hours.\footnote{We set a fixed time limit of three hours based on \citet{mirowski2023cowriting} who allowed $2$ hours to participants for co-writing a screenplay.} We do not enforce a minimum number of interactions on the participants, but encourage them to explore model capabilities in order to give us feedback.\footnote{We note that our recruitment procedure selects writers who were interested in exploring the possibility of human-AI collaborative writing, potentially resulting in a sample of participants naturally inclined to be more curious about the model capabilities.} We did not ask participants to complete the writing session in one block but instead allowed them to break the process in whatever time ratio they felt comfortable with. Once they submit their story, we ask them to complete a feedback form, as discussed in \Cref{sec:writer_feedback}. Depending on their availability, writers had the option of contributing up to $3$ stories. For remuneration, we asked participants to sign up through \textit{Upwork}\footnote{\url{https://www.upwork.com/}} (a professional freelancing website) and paid them a total of \$$90$ (\$$30$/hr) for each story. We also sent a bonus payment of \$$50$ to every participant who wrote $2$ stories as a token of their commitment to the study.
\section{Investigating User Study Data}
\label{sec:interaction}

\subsection{Collected Stories}
\begin{figure*}
    \subfloat[Histogram of word counts of all the stories written by our participants]{\includegraphics[width=0.45\textwidth]{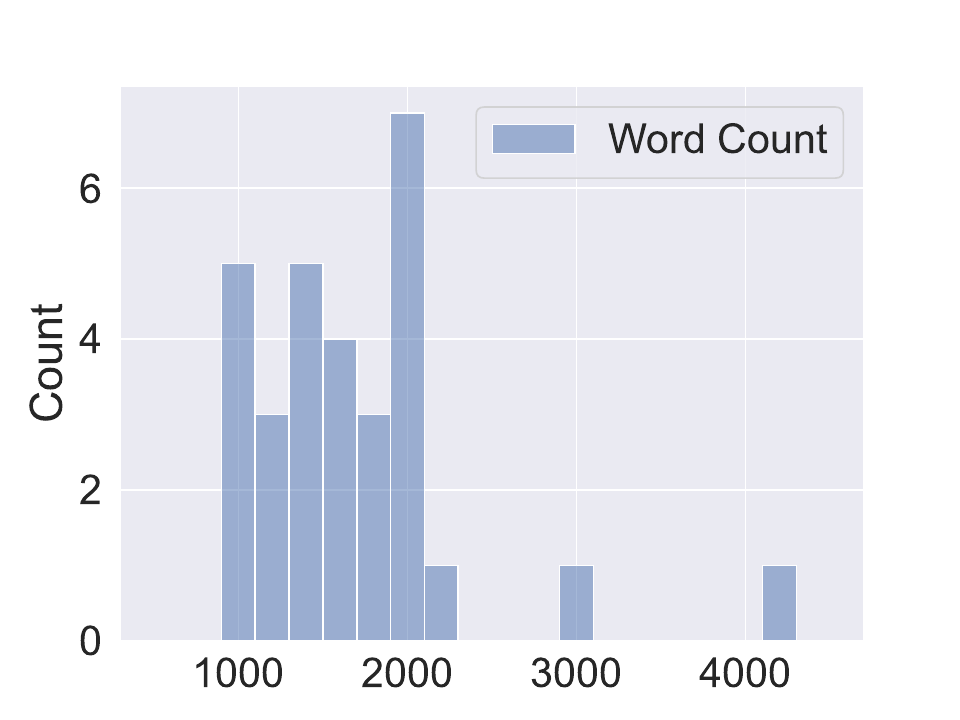}} %\quad
    \hfill
    \subfloat[Distribution of Genre across all the stories written by our participants]{%\frame{
    \includegraphics[width=0.48\textwidth]{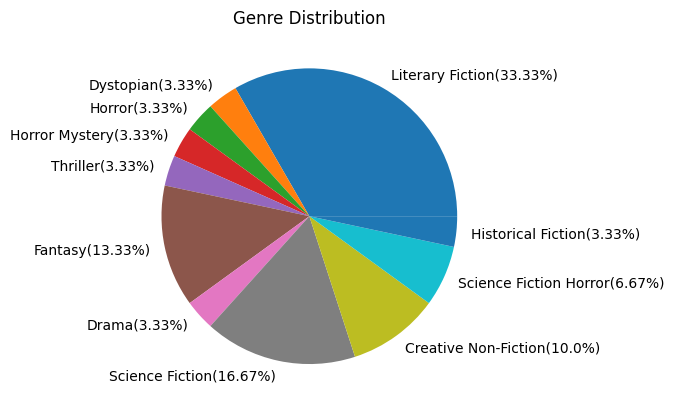}}
    %} 
    \caption{\label{statistics} Summary Statistics on Stories}
\end{figure*}

\begin{table}[!ht]
\centering
\small
\begin{tabular}{|l|l|l|l|l|}
\hline
Index & Title                                      & Author & Genre         & Theme                                                                                                                                         \\ \hline
1 & Lesbian Leverage (2980)  & MG   & Literary Fiction  & Identity, Self Discovery \& Power Dynamics  \\ \hline %Shrinking Scholar (2010)  & LHS     & Science Fiction Horror. & \begin{tabular}[c]{@{}l@{}}Cosmic terror and madness in the face of the \\incomprehensible. \end{tabular}   \\ \hline
2 & Inheritance of Shadows (4290)                  & MG   & Horror Mystery    & \begin{tabular}[c]{@{}l@{}}Suffering \& abuse caused by cruel mothers \\\& the quest for justice \& freedom from them.\end{tabular} \\ \hline %Words in the Wilderness (1322)            & MC        & Literary Fiction  & Self-delusion vs Self-growth.  \\ \hline
3 & \begin{tabular}[c]{@{}l@{}}Between Smoke Breaks and \\Flickering Dreams (1206)\end{tabular} & SH & Literary Fiction  & Hope, Struggle and Disillusionment  \\ \hline %Nostalgia and Early AI (1025)           & ATF          & Creative Non-Fiction       & \begin{tabular}[c]{@{}l@{}}Human's Relationship \& Connection\\ with Technology.\end{tabular} \\ \hline
4 &  Preserved Fears (1523)                          & SH  & Dystopian         & \begin{tabular}[c]{@{}l@{}}Strained relationship between a mother \\suffering from paranoid delusions \&\\ her concerned daughter\end{tabular} \\ \hline
5 & Infernal Melody (981)  & SH  & Horror   & Teenage Naivety \\ \hline %Galactic Rainbow(1984)                    & Anon       & Science Fiction   & Acceptance \& Identity.  \\ \hline
6 & This eve of parting (1579)  & MC    & Literary Fiction  & Memories \& mortality \\ \hline %The Party (1355)                           & ES     & Literary Fiction  & Healing \& finding closure. \\ \hline
7 & Words in the Wilderness (1322)            & MC        & Literary Fiction  & Self-delusion vs Self-growth.  \\ \hline %Gator Land (1333)                     & KK          & Science Fiction Horror   & Obsession \& its dangers. \\ \hline
8 & Revenge Game (2172)  & MC        & Thriller    & Vengeance \\ \hline %\begin{tabular}[c]{@{}l@{}}A chair falls in\\love with a machine (1297)\end{tabular} & REB     & Science Fiction. & Resilience in the face of loss \& change. \\ \hline
9 & The Broken Clock (1751)  & ES       & Fantasy           & Fleeting nature of time. \\ \hline %Dyed in the Wool (1923)  & ES     & Drama             & Acceptance \& Inclusiveness.  \\ \hline
10 & Dyed in the Wool (1923)  & ES     & Drama             & Acceptance \& Inclusiveness.  \\ \hline %\begin{tabular}[c]{@{}l@{}}Neo Metropolis \& \\the Silver Delta (2006)\end{tabular}    & Anon     & Science Fiction   & Love, courage \& Transformation.  \\ \hline
11 & The Party (1355)                           & ES     & Literary Fiction  & Healing \& finding closure. \\ \hline %Clean Revelations (1493)                   & ATK      & Literary Fiction. & Appreciation of ordinary things in life.  \\ \hline
12 & Nostalgia and Early AI (1025)           & ATF          & Creative Non-Fiction       & \begin{tabular}[c]{@{}l@{}}Human's Relationship \& Connection\\ with Technology.\end{tabular} \\ \hline %The Brine Pool Dream (1217)  &  FV   &  Fantasy  & Imagination \& escapism \\ \hline
13 & Mirror Morsels: Body, Pet, and Self (995)   & ATF     & Creative Non-Fiction & Body image and Self Acceptance.  \\ \hline % Infernal Melody (981)  & SH  & Horror   & Teenage Naivety \\ \hline
14 & Bread and Butter: A Service Journey (1992)    & ATF     & Creative Non-Fiction       & \begin{tabular}[c]{@{}l@{}}Investing in employees: key\\ to restaurant success.\end{tabular}  \\ \hline %The Limp (902)  &    NM   & Fantasy & \begin{tabular}[c]{@{}l@{}}Suffering \& avoidance.\end{tabular} \\ \hline
15 & Gator Land (1333)                     & KK          & Science Fiction Horror   & Obsession \& its dangers. \\ \hline %Lesbian Leverage (2980)  & MG   & Literary Fiction  & Identity, Self Discovery \& Power Dynamics  \\ \hline
16 & \begin{tabular}[c]{@{}l@{}}A chair falls in\\love with a machine (1297)\end{tabular} & REB     & Science Fiction. & Resilience in the face of loss \& change. \\ \hline %Artful Detour  (1411)  & NM    & Literary Fiction & \begin{tabular}[c]{@{}l@{}}Friendship, relationships \& overcoming\\ differences.\end{tabular} \\ \hline
17 & Writing with the Machines (2079)           & CK      & Science Fiction   & \begin{tabular}[c]{@{}l@{}}The perils of artificial intelligence and \\technology gone awry.\end{tabular} \\ \hline %The Broken Clock (1751)  & ES       & Fantasy           & Fleeting nature of time. \\ \hline
18 & Clean Revelations (1493)                   & ATK      & Literary Fiction. & Appreciation of ordinary things in life.  \\ \hline %\begin{tabular}[c]{@{}l@{}}The Work of Art in the Age\\ of Mechanical Reproduction (1521)\end{tabular}  & GS   &   Historical Fiction &  \begin{tabular}[c]{@{}l@{}}Facing injustice, an intellectual defiantly \\leaves behind an enduring legacy.\end{tabular} \\ \hline
19 & Artful Detour  (1411)  & NM    & Literary Fiction & \begin{tabular}[c]{@{}l@{}}Friendship, relationships \& overcoming\\ differences.\end{tabular} \\ \hline %The Clown of God (904)  & AF & Fantasy &  \begin{tabular}[c]{@{}l@{}}Search for meaning \& connection in a harsh,\\ exploitative world\end{tabular} \\ \hline
20 & The Limp (902)  &    NM   & Fantasy & \begin{tabular}[c]{@{}l@{}}Suffering \& avoidance.\end{tabular} \\ \hline %Inheritance of Shadows (4290)                  & MG   & Horror Mystery    & \begin{tabular}[c]{@{}l@{}}Suffering \& abuse caused by cruel mothers \\\& the quest for justice \& freedom from them.\end{tabular} \\ \hline
21 &  Shrinking Scholar (2010)  & LHS     & Science Fiction Horror. & \begin{tabular}[c]{@{}l@{}}Cosmic terror and madness in the face of the \\incomprehensible. \end{tabular}   \\ \hline %Bread and Butter: A Service Journey (1992)    & ATF     & Creative Non-Fiction       & \begin{tabular}[c]{@{}l@{}}Investing in employees: key\\ to restaurant success.\end{tabular}  \\ \hline
22 & \begin{tabular}[c]{@{}l@{}}I know it's not the same,\\ but it's something (1736) \end{tabular}  &  MK  & Literary Fiction &   \begin{tabular}[c]{@{}l@{}}Technology as a means to process grief  \end{tabular} \\ \hline %The Pursuit of Perfection (2086)          & Anon        & Literary Fiction  & Pursuit of Perfection \& its perils.  \\ \hline
23 & The Brine Pool Dream (1217)  &  FV   &  Fantasy  & Imagination \& escapism \\ \hline %Ill Repute (1771) & RH   &  Science Fiction &  \begin{tabular}[c]{@{}l@{}}Technology's dual nature; the peril of\\ misinformation; \& the possibility of\\ redemption.\end{tabular}  \\ \hline
                                
24 & s  l  o  w  (1592)  &  FV     & Literary Fiction &   Identity \& self-acceptance. \\ \hline % Revenge Game (2172)  & MC        & Thriller    & Vengeance \\ \hline
25 & \begin{tabular}[c]{@{}l@{}}The Work of Art in the Age\\ of Mechanical Reproduction (1521)\end{tabular}  & GS   &   Historical Fiction &  \begin{tabular}[c]{@{}l@{}}Facing injustice, an intellectual defiantly \\leaves behind an enduring legacy.\end{tabular} \\ \hline %This eve of parting (1579)  & MC    & Literary Fiction  & Memories \& mortality \\ \hline
26 & The Clown of God (904)  & AF & Fantasy &  \begin{tabular}[c]{@{}l@{}}Search for meaning \& connection in a harsh,\\ exploitative world\end{tabular} \\ \hline %\begin{tabular}[c]{@{}l@{}}Between Smoke Breaks and \\Flickering Dreams (1206)\end{tabular} & SH & Literary Fiction  & Hope, Struggle and Disillusionment  \\ \hline
27 & Ill Repute (1771) & RH   &  Science Fiction &  \begin{tabular}[c]{@{}l@{}}Technology's dual nature; the peril of\\ misinformation; \& the possibility of\\ redemption.\end{tabular}  \\ \hline %\begin{tabular}[c]{@{}l@{}}I know it's not the same,\\ but it's something (1736) \end{tabular}  &  MK  & Literary Fiction &   \begin{tabular}[c]{@{}l@{}}Technology as a means to process grief  \end{tabular} \\ \hline
28 & \begin{tabular}[c]{@{}l@{}}Neo Metropolis \& \\the Silver Delta (2006)\end{tabular}    & Anon     & Science Fiction   & Love, courage \& Transformation.  \\ \hline %Mirror Morsels: Body, Pet, and Self (995)   & ATF     & Creative Non-Fiction & Body image and Self Acceptance.  \\ \hline
29 &  Galactic Rainbow(1984)                    & Anon       & Science Fiction   & Acceptance \& Identity.  \\ \hline %Writing with the Machines (2079)           & CK      & Science Fiction   & \begin{tabular}[c]{@{}l@{}}The perils of artificial intelligence and \\technology gone awry.\end{tabular} \\ \hline
30 & The Pursuit of Perfection (2086)          & Anon        & Literary Fiction  & Pursuit of Perfection \& its perils.  \\ \hline %s  l  o  w  (1592)  &  FV     & Literary Fiction &   Identity \& self-acceptance. \\ \hline
\end{tabular}
\vspace{2ex}
\caption{\label{data}30 stories written by 17 MFA students across varying genres and themes. The number between parenthesis in the title column show the word length %\sm{Group this by authors, W1-W17. Also maybe even in the orginal author table put first the ones with 3 stories, then 2 and then 1. And then the story intex in order of writers.}
}
\end{table}

We collected $30$ stories from participants between February and July 2023, detailed in Table \ref{data} and Figure \ref{statistics}. Story lengths varied mostly between %, with most between 
$1000$ and $2000$ words with two outliers on the longer side.%although some exceeded this limit. For example, 
\footnote{Both longer stories were contributed by MG, one of which was a 4290-word horror mystery %story written with our interface 
that she broke into three separate acts to deal with the token limit of GPT-3.5.} %Overall most stories ranged between 1000 and 2000 words.
Our study participants wrote stories across $11$ genres, with \textit{Literary Fiction} being the most preferred. Some blended genres, like KK and LHS's \textit{Science Fiction} and \textit{Horror} mix. Others, like MG and GS, wrote in \textit{Horror Mystery} and \textit{Historical Fiction}, respectively. We asked participants to summarize their story's central theme in $5$-$15$ words. As Table \ref{data} shows, themes ranged from \textit{Body Image, Acceptance, Identity, Self Discovery} as well as \textit{Relationships} or \textit{Impacts of Technology}. Appendix \ref{app:story_example} contains an example of a story plot written by the user, model initial draft, user interactions, and the final story submitted as part of the user study.

Our work aims to understand how expert users utilize LLMs during story writing to identify common concerns as future directions of exploration. In line with recommendations from \cite{shen2023parachute, lee2022evaluating}, we look for patterns both in the interaction between expert writers and LLM (\Cref{sec:interaction_patterns}) as well as the model contribution on the final writing artifact \ie the stories collected from our study (\Cref{sec:model_contribution}). The first aims to understand for which activities writers seek the help of LLMs during the writing process (e.g., planning, translation, and reviewing), while the latter aims to understand in which of these activities the LLM actually helps. 

\label{sec:data_collection}
% \label{sec:interface}

\subsection{Identifying Patterns in User Interactions}
\label{sec:interaction_patterns}

\begin{figure}[!htp] 
    \small
    \centering
    \begin{subfigure}[b]{0.49\textwidth}
        \centering
        \includegraphics[width=\textwidth]{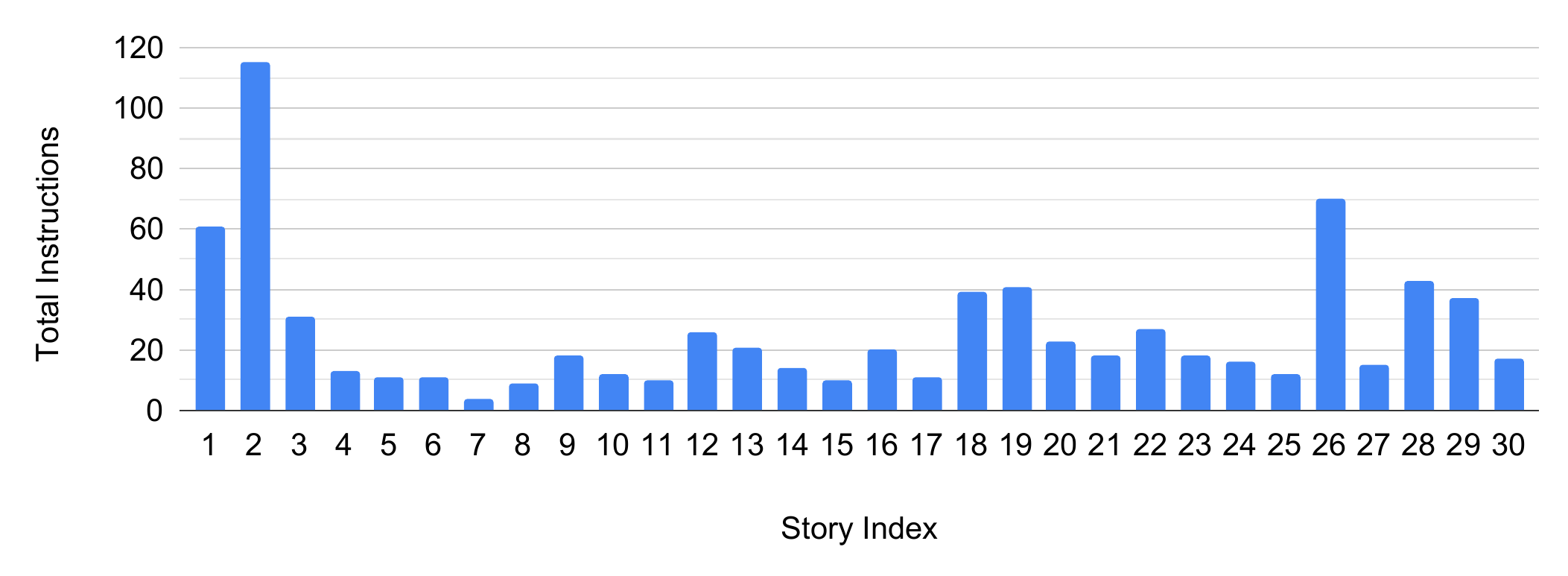}
    \caption{Total instructions ($y$-axis) requested by the writers for each story ($x$-axis)}
        \label{fig:tot_inst}
    \end{subfigure}
    \hfill
    \begin{subfigure}[b]{0.49\textwidth}
        \centering
        \includegraphics[width=\textwidth]{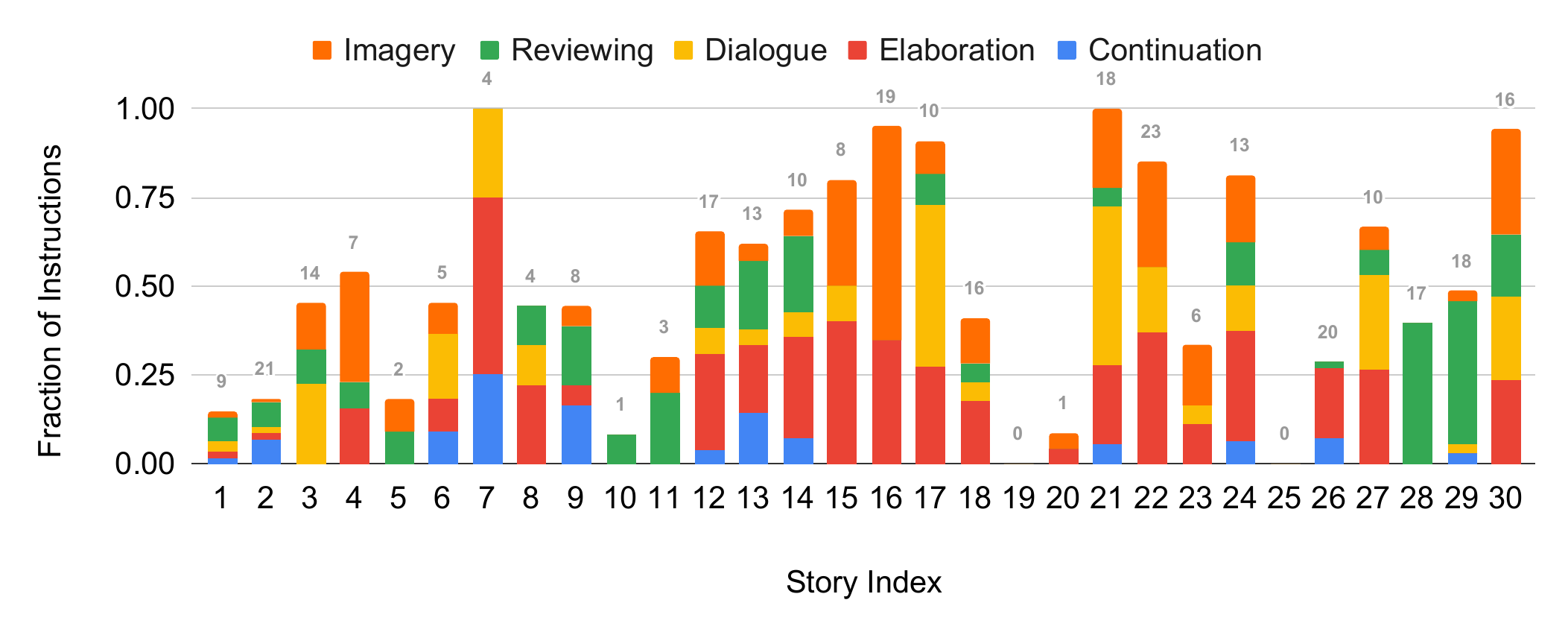}
        \caption{Fraction of templated instructions ($y$-axis) along with their total count (above each bar) per story.}
        \label{fig:templated_figure}
    \end{subfigure}
    \caption{Summary statistics on user interactions showing (a) the total count of instructions asked per user and (b) the fraction, and total count, of these which used the templates that we provided (\Cref{sec:draft_chat})}
    \label{fig:instruction_counts}
\end{figure}

\paragraph{\textbf{1. Frequency of human-LLM interactions}}
First, we examine the total number of instructions asked by each user, as shown in \Cref{fig:tot_inst}. The instruction counts display significant variance, ranging from $4$ to $115$, with a median of $18$ and a mean of $25.76$. Excluding outliers, $21$ out of $30$ stories included between 10 and 30 instructions indicating substantial collaboration between the humans and models.  The outlier values can be categorized into two groups. The first group consists of a pair of stories, both by MC, who requested less than 10 instructions.%---possibly due to user preferences rather than explicit flaws in the user interface or model. 
\footnote{In the post-completion feedback form, MC indicated that they maintained a high level of agency in writing the majority of the story compared to the model, finding the suggestions only somewhat helpful (\Cref{sec:writer_feedback}).} %\sm{So if stories are grouped by writer this can be immediately seen. Seems to me this writer is MC, can you check how he considered helpfulness of the LLM (in Fig 7 that will be helpful to know)}
The second group includes five stories with over $40$ instructions showcasing a desire from our participants to engage with the model and understand its capabilities within their writing process. In four of these five stories, the writers reported finding the model to be helpful (\Cref{sec:writer_feedback}). % This underscores their strong desire to engage with the model as they understand its competencies to use them effectively.

%\begin{figure}[!ht]
%   \centering
%    \small
%   \includegraphics[width=0.7\textwidth]{figures/Total_Instructions_vs._Story_Index.pdf}
%    \caption{Total instructions requested by the writers for each story}
%    \label{fig:tot_inst}
%\end{figure}

Second, we note that our user interface presents users with templated instructions (\Cref{sec:draft_chat}). % that were created following input from other writers and prior work \cite{yuan2022wordcraft}. To see how this primes the usage of the model, we calculate the fraction of instructions requested that follow one of these templates using regex matching (\Cref{fig:templated_figure}).
We observe that the total instruction count and fraction of templated instructions have a Pearson correlation coefficient of $-0.37$ indicating that users who ask for more instructions tend to use fewer templated ones, instead opting for instructions with a higher level of specificity (\Cref{fig:templated_figure}).\footnote{This result is significant at the $5\%$ level with a p-value for the correlation of $0.04$.} We provide examples of templated instructions asked by the writers in \Cref{tab:promptinsp}.

\begin{table}[!ht]
    \def\arraystretch{1.15}
    \centering
    \small
    \begin{tabular}{|ll|}
        \hline
        \multicolumn{2}{|l|}{{\color{ao}Rewrite the following text with more rich imagery}: \textit{The rooms were cozy, the food hearty.......}}  \\ \hline
        \multicolumn{1}{|l|}{SH}  & \begin{tabular}[c]{@{}l@{}}Rewrite the following text with more rich and less cliche imagery, taking into account the tone and aesthetics of the \\piece so far: \textit{People were breathing altogether .....}\end{tabular}    \\ \hline
        \multicolumn{1}{|l|}{RH}  & \begin{tabular}[c]{@{}l@{}}Rewrite the following text so that it better portrays the characters' emotional turmoil and the gradual deterioration of\\ their judgment
        and sanity: \textit{Laura's voice quivered with frustration......}\end{tabular} \\ \hline
        \multicolumn{1}{|l|}{ATF} & Rewrite the following text with more imagery like Mary Shelley's in Frankenstein: ....         \\ \hline\hline
        \multicolumn{2}{|l|}{{\color{ao}Elaborate the following text. Add more details}: And he told them that he had once blackmailed a CEO}  \\ \hline
        \multicolumn{1}{|l|}{ATF} & \begin{tabular}[c]{@{}l@{}}Elaborate the following text by one sentence. Add a transition to the rest of the story draft: \textit{It may not have been} \\\textit{on a dreary night of November....}\end{tabular}   \\ \hline
        \multicolumn{1}{|l|}{RH}  & \begin{tabular}[c]{@{}l@{}}Elaborate the following text. Add more details in which Laura and James take more precautions \& thoroughly investigate\\ the Oracle's credibility before trusting them: \textit{In their a quest to end the reign their deep fake avatars, Laura and James....}\end{tabular} \\ \hline
    \end{tabular}
    \vspace{2ex}
    \caption{Example prompts from different users inspired from our templated prompts}
    \label{tab:promptinsp}
\end{table}

\begin{table}[!ht]
    \centering
    \small
    \def\arraystretch{1.1}
    \begin{tabular}{|l|l|l|l|}
        \hline
        ATF & Nostalgia \& Early AI & \begin{tabular}[c]{@{}l@{}}Given the story write some critiques or inconsistencies citing portions. Would this\\ story work better with the example of Ask Jeeves? Give your reasons\end{tabular}  & Reviewing   \\ \hline
        SH  & Preserved Fears        & \begin{tabular}[c]{@{}l@{}}Do you think this story would be better suited written in the 1st person point of \\ view of Jude?\end{tabular} & Reviewing   \\ \hline
        SH  & Preserved Fears        & \begin{tabular}[c]{@{}l@{}}I have deleted the end of this story as I want it to take place over the course of a \\ single conversation and I also don't want Ellen to be correct about the end of the \\ world, she is supposed to be loving but delusional. What are some possible \\ directions for this conversation to go in so that there is a successful story arc?\end{tabular}      & Planning    \\ \hline
        Anon & Galactic Rainbow       & Write a thematically relevant reason why the earth exploded   & Planning    \\ \hline
        ES  & The Party   & \begin{tabular}[c]{@{}l@{}}Rewrite this passage to give Addie and Becca more personality, and give us an \\ idea of their relationship through their manner of speech and the way they greet \\ each other. Do not use explanatory language, you must "show" me all this through\\ a scene: {\color{ao}She gave Addie a quick rundown of the night's events. "Let's start with} \\ {\color{ao}something easy, like finding you a drink and introducing you to everyone."}\end{tabular} & Translation \\ \hline
        ES  & Dyed in the Wool       & \begin{tabular}[c]{@{}l@{}}Given your feedback, can you make recommendations of examples of how the \\ the story might be improved; for example, by creating characters in the church \\ who might oppose David \& Sarah?\end{tabular}        & Planning    \\ \hline
        ES  & Dyed in the Wool       & \begin{tabular}[c]{@{}l@{}}Try writing the scene where Sarah reveals she is transgender to David, avoid \\ cliches and obvious exchanges of dialogue. Use examples like Little Miss \\ Sunshine to make them quirky and mid-western in their speech and mannerisms\end{tabular}& Translation \\ \hline
        ATK & Clean Revelations      & \begin{tabular}[c]{@{}l@{}}Can you write an exchange between Martin and the laundromat attendant in the \\ style of Flannery O'Connor? Can the attendant extoll the virtues of mechanization\\ to Martin?\end{tabular}  & Planning    \\ \hline
        MG & \begin{tabular}[c]{@{}l@{}}Inheritance of Shadows\end{tabular}       & \begin{tabular}[c]{@{}l@{}}Can you think of a twist ending for this story instead of the two endings\\ I do have in each section of this short story?
        \end{tabular}  & Planning   \\ \hline
        FV & The Brine Pool Dream       & \begin{tabular}[c]{@{}l@{}}Make Lo less pragmatic and more confused\end{tabular}  & Translation    \\ \hline
        NM & The Limp       & \begin{tabular}[c]{@{}l@{}}Could you rewrite the first paragraph as Tolstoy would? That is to say, elegantly\\ and driving straight into the action
        \end{tabular}  & Translation    \\ \hline
        NM & Artful Detour & \begin{tabular}[c]{@{}l@{}}All the paragraphs in the story begin with very predictable words: `The,' `at first,' \\`it all began.' Could you make it more interesting?\end{tabular}  & Translation    \\ \hline
        MG & Lesbian Leverage       & \begin{tabular}[c]{@{}l@{}}Show me an example of where you think Holly's thoughts and feelings \\should be condensed
        \end{tabular}  & Reviewing    \\ \hline
    \end{tabular}
    \vspace{2em}
    \caption{ Examples of non-templated instructions for Planning, Translation, and Reviewing that are written by our participants}
    \label{tab:nontemplated}
\end{table}

 \begin{figure}
    \centering
     \includegraphics[width=0.7\textwidth]{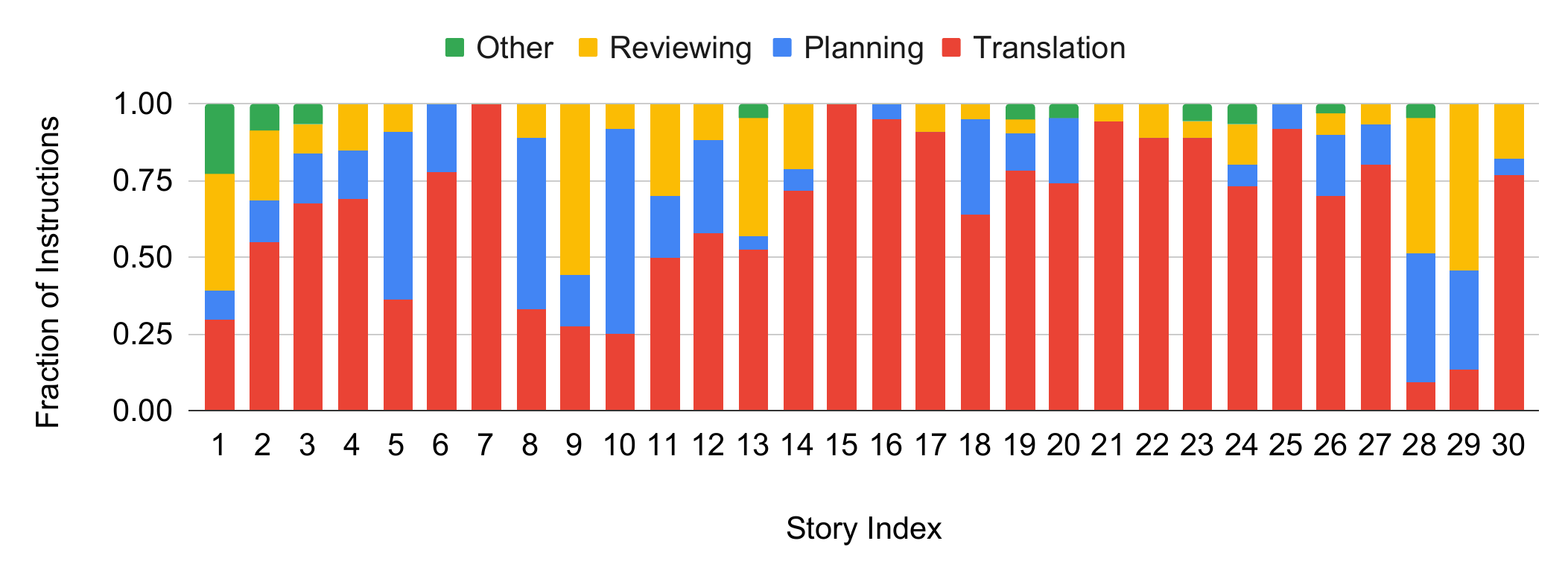}
     \caption{Fraction of instructions corresponding to each cognitive activity for each story}
     \label{fig:category_figure}
     %\end{subfigure}
    %\caption{}
    \label{fig:categorizing_instructions}
\end{figure}

\begin{figure*}
    \subfloat[Inheritance of Shadows]{\includegraphics[width=0.8\textwidth]{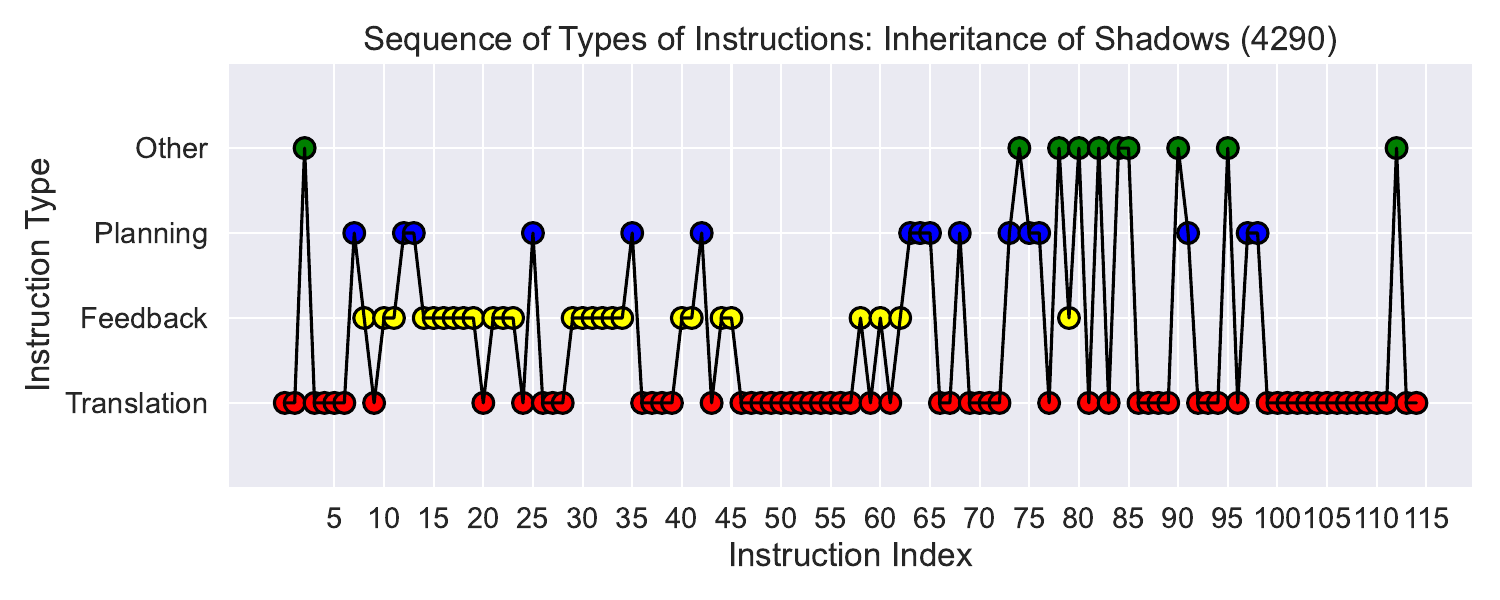}} \\
    \subfloat[Galactic Rainbow]{\includegraphics[width=0.8\textwidth]{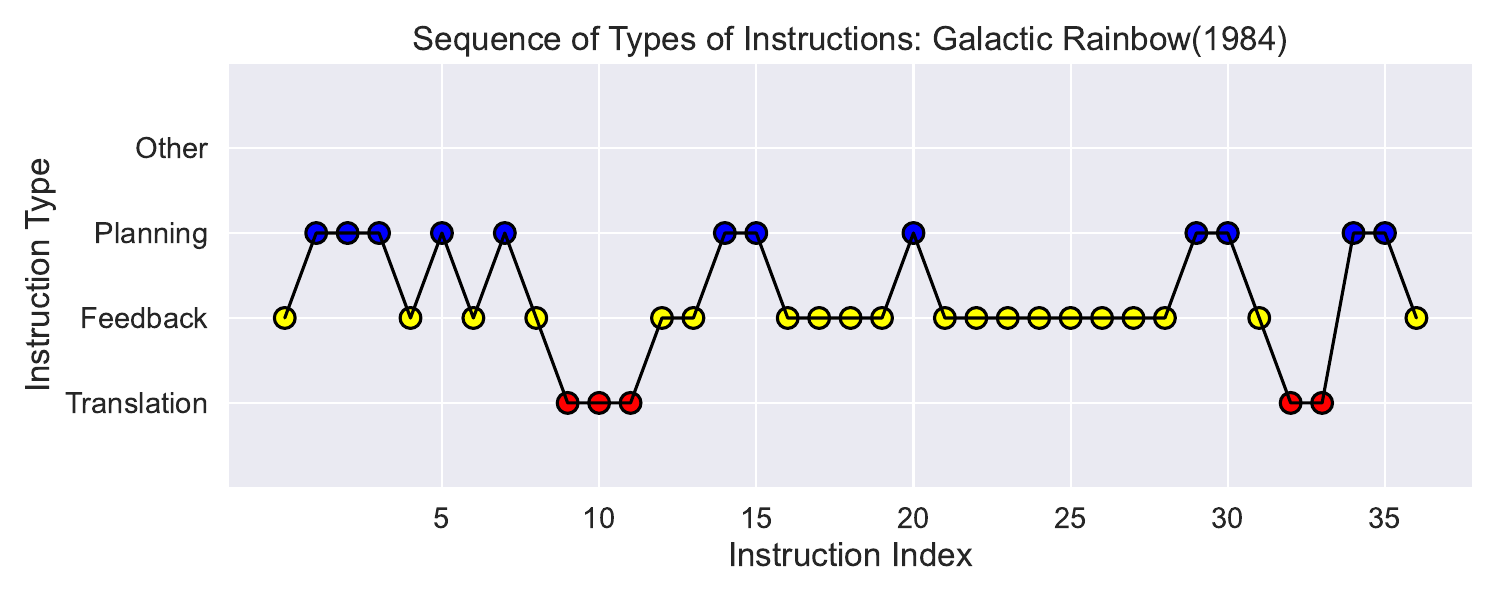}} 
    \caption{\label{fig:sequence_inst} Writers employ a non-linear writing process, alternating between planning, reviewing, and translation-based instructions when interacting with the model. This shows the value in our design choice to separate the draft writing and model interactions (\Cref{sec:draft_chat}). }
\end{figure*}

\paragraph{\textbf{2. Categorizing the intent of user instructions}}
We also note that the users seek model support for various kinds of tasks in the writing process which are mainly categorized into planning, reviewing, and translation \cite{gero2023social,gero-etal-2022-design}. To better understand the kind of assistance our writers seek from the model, we manually annotate the collected instructions into the three proposed categories. 
Each instruction was independently annotated thrice by the authors of this work and the label was determined via a majority vote.\footnote{The Fleiss' kappa of annotations was $0.52$, indicative of moderate agreement on the task.} \Cref{fig:category_figure} illustrates that the majority of instructions fall under the translation label, involving local editing subtasks such as introducing imagery or making targeted changes to the story draft.
Moreover, a relatively large fraction of suggestions pertain to seeking feedback from the model on the draft, suggesting a potential avenue for models to offer intermediate feedback, helping writers overcome mental blocks and improve their drafts. Qualitative feedback from our users also indicates that these are more closely aligned with the model's strengths (\Cref{sec:writer_feedback}). We also observe relatively similar usage patterns when the same writer contributes multiple stories---story indices 1 and 2 by MC and indices 9, 10, and 11 by ES to name two.  Writers also ask a wide variety of planning-based instructions that tend to be highly specific to the story at hand. Examples include introducing dialogue between characters incorporating their traits as inferred from the draft and collectively brainstorming alternative endings to the story (\Cref{tab:nontemplated}). However, users find the model output lacking on these, as further discussed in \Cref{sec:writer_feedback} and \Cref{sec:discussion}. 

\paragraph{\textbf{3. Illustrating the non-linearity of the writing process via observed user instructions}} Annotating the user instructions for intent also allows us to observe the non-linearity of the writing process employed by our users. We plot the intents of the instructions in sequence for two stories in \Cref{fig:sequence_inst}. We observe that both authors alternate between planning, translation, and reviewing-based instructions. This shows a visual represention of the cognitive process model of writing \cite{flower1981cognitive} in practice and highlights the value of our theoretically grounded interface.  
We see how separating the interface for draft-writing and interacting with the model (\Cref{sec:draft_chat}) better suits the needs of the writers. 

\paragraph{\textbf{4. Are writers good prompt engineers?}}
An important consideration for our work is that the expert writers often have precise and unique instructions for the model during the writing process and high standards for the text they are willing to integrate into the story.  %(\Cref{sec:user_recruiting})\fb{$\leftarrow$ do we need this reference?}. 
Recent work has shown that prompting the model suitably in order to elicit these desired behaviors can be challenging for non-AI experts \cite{10.1145/3544548.3581388}. Our design lets users interact with the model maintaining the chat history so they can observe the output text from the model and reframe their request iteratively if needed. %ES notes in their feedback \emph{``Its creativity also depends on my ability to be more specific in my prompts, so the more I give it the better job it can do.''}
Based on the interaction data, we observed several participants were successfully able to design expressive and high-quality prompts. Some of these prompt designs were inspired by our provided templated prompts as can be seen in \Cref{tab:promptinsp}. Additionally, we noticed that several participants came up %were innovative and effective at coming up 
with novel prompts to steer the story-writing process. \Cref{tab:nontemplated} shows some examples of well-engineered non-templated prompts. % \fb{maybe define what's a well-engineered prompt? are the prompts for which model outputs satisfying answers?}. 

\paragraph{\textbf{5. How do Genre writers use LLMs for creativity support?}}

\begin{table}[!ht]
    \def\arraystretch{1.15}
    \centering
    \small
    \begin{tabular}{|l|p{1.95cm}|l|}
        \hline
        KK & SciFi Horror & Can you write this in the style of Cormac Mccarthy    \\ \hline
        Anon & Science Fiction        & Short descriptive world-building passage about space in the story\\ \hline
        Anon & Science Fiction        & Write a new name for a post-modern steampunk dystopian New York City        \\ \hline
        NM & Fantasy & Come up with a fictitious name for a Swiss hotel \\ \hline
        GS & Historical Fiction     & \begin{tabular}[c]{@{}l@{}}Write the scene in which Walter Benjamin kills himself with morphine tablets in a spare room at the\\ French-German border in 1940 while escaping from the Nazis. Make unusual observations\end{tabular} \\ \hline
        MC & Thriller    & Write a Hollywood action-style revenge fantasy against the owner of the Knicks \\ \hline
        MG & Horror Mystery         & What are plants that would be in a haunted garden?    \\ \hline
        MG & Horror Mystery         & \begin{tabular}[c]{@{}l@{}}Write five scariest sentences you can think of related to this short story in the literary horror genre.\end{tabular}\\ \hline
        MG & Horror Mystery         & Re-write it but make it much more scary. focus on psychological horror aspects.        \\ \hline
    \end{tabular}
    \vspace{2ex}
    \caption{Instructions pertaining to genre-specific writing}
    \label{tab:genre}
    \vspace{-4ex}
\end{table}
Due to the diverse nature of the stories written by our users, we also find 
that participants who chose to write in a particular genre often used the model to either adhere to the specific style representing that genre or used it for \textit{World Building}. World-building often refers to the creation of a world that is different from our own. We can see this most obviously in genres like fantasy, horror, or science-fiction, where the writer has created something new and exciting for the reader to delve into. %Similarly, spy thrillers although usually set in our known world, are also doing a heavy amount of world-building. As can be seen in 
As seen in \Cref{tab:genre}, KK desired for the model to adapt to the style of \textit{Cormac Mcarthy} who is famously known for his short stories spanning the Western and post-apocalyptic genres. Similarly, MG expected the model to know and adapt to the specifics of the psychological horror genre. Other examples show how participants use the model for world-building either by instructing to draft a genre-specific scene (\Cref{tab:genre}  Row 2,5,6) or simply suggestions of characters or elements that make the world believable at the sensory level (\Cref{tab:genre} Row 3,4,7).

\subsection{Model Contribution to the Final Story}
\label{sec:model_contribution}
\begin{figure}[!htp]
    \small
    \begin{subfigure}[b]{0.475\textwidth}
        \centering
         \includegraphics[width=\textwidth]{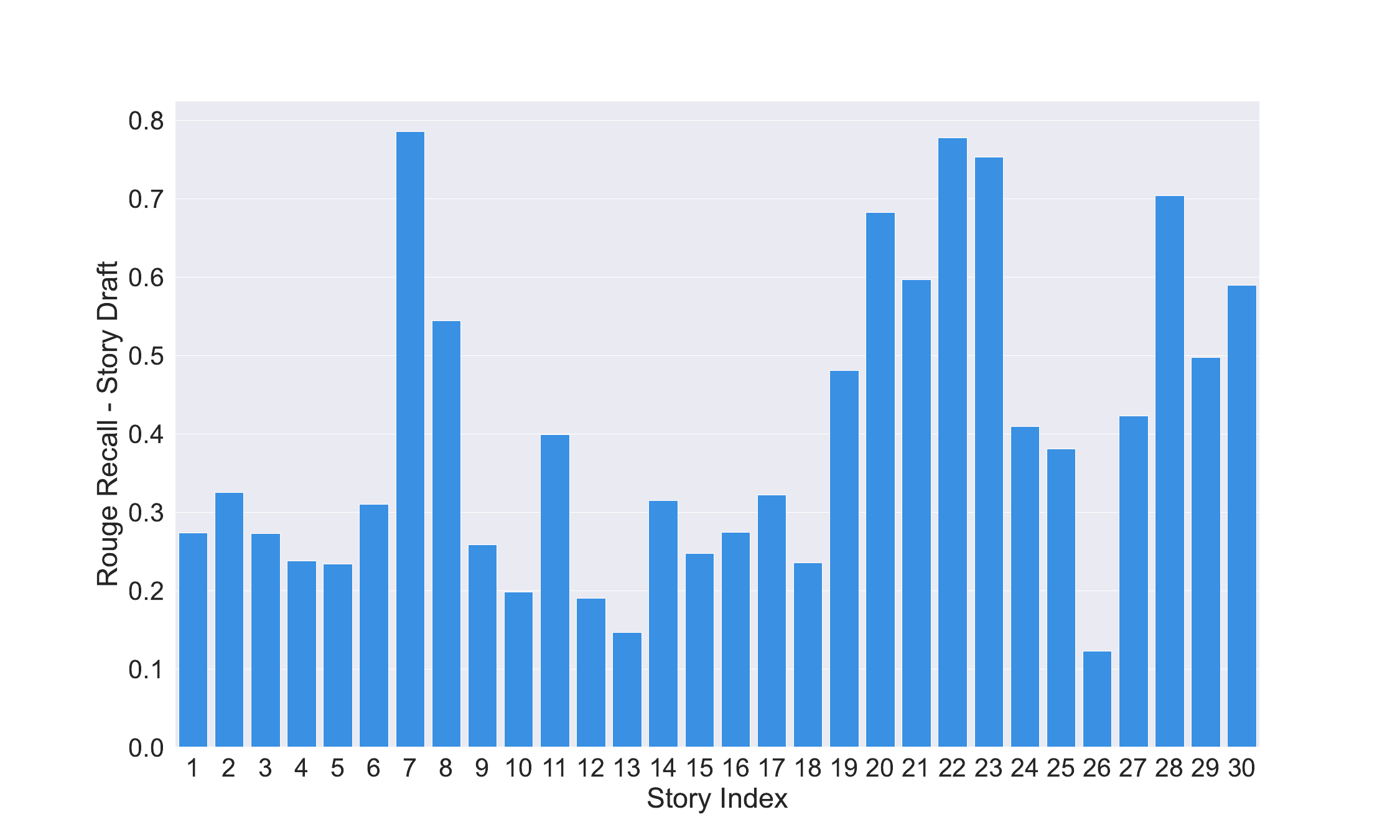}
         \caption{Fraction of the initial model-generated story draft included in the final story submitted by the writer ($y$-axis) for various stories ($x$-axis) in the user study. We calculate this fraction using Rouge-L recall \cite{lin-2004-rouge}.}
         \label{fig:rougeplot_draft}
     \end{subfigure}
     \hfill
     \begin{subfigure}[b]{0.475\textwidth}
        \centering
         \includegraphics[width=\textwidth]{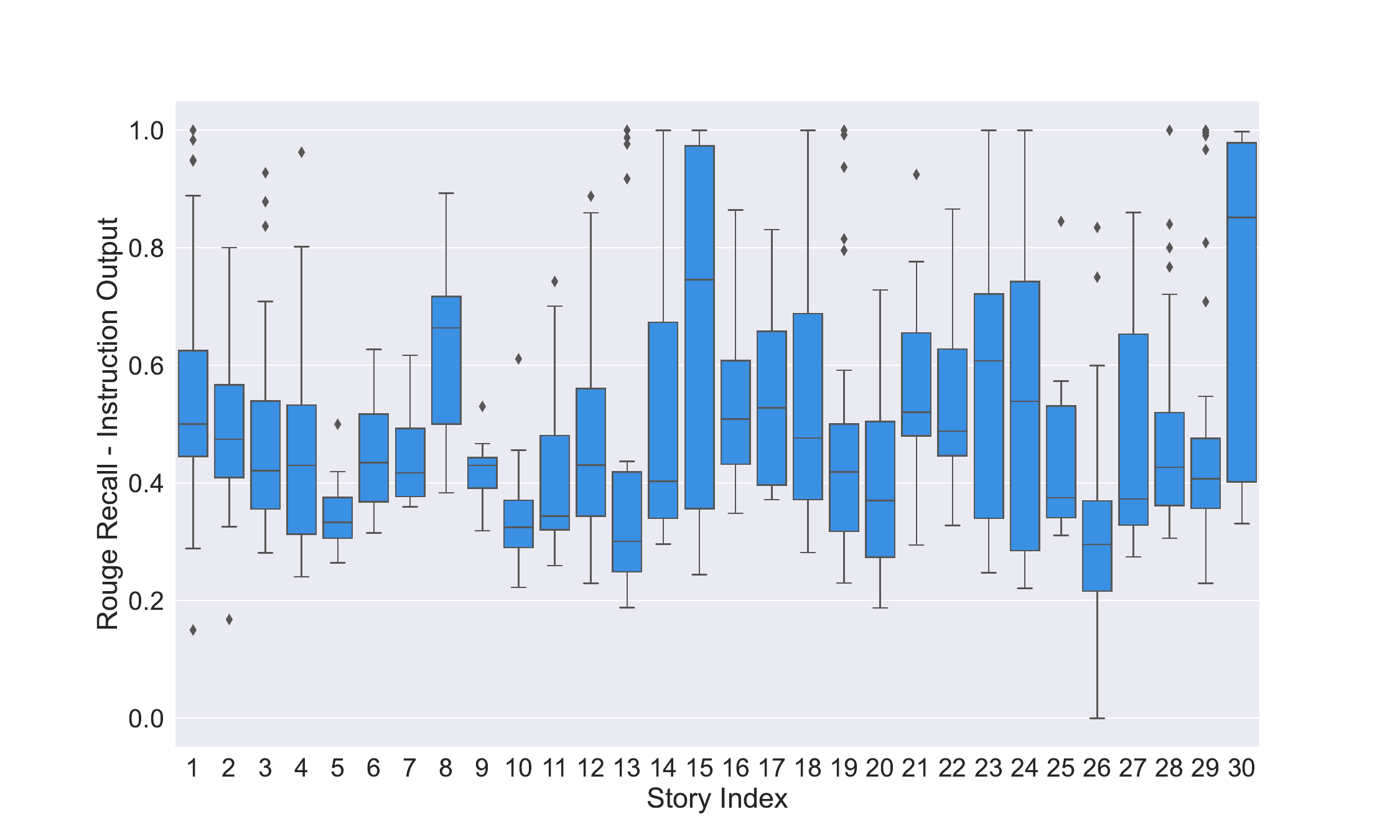}
         \caption{Fraction of the model-generated text in response to different user instructions that was retained in the final story submitted by the writer ($y$-axis) for various stories ($x$-axis) in the user study. We calculate this fraction using Rouge-L recall \cite{lin-2004-rouge}.}
         \label{fig:rougeplot_instructions}
     \end{subfigure}
    \caption{}
    \label{fig:rougeplots}
\end{figure}

%\sm{Maybe call it "Understanding where LLMs helped" or something like that. this points back to the aims}
In addition to examining user interactions to understand where do writers seek the help of LLMs (planning, translation, reviewing), we wish to quantitatively estimate the model influence on the the stories collected in our study. Our interface enables users to utilize the model to compose an initial draft, which they subsequently refine (\Cref{sec:plot_to_draft}). The fraction of this model written draft which is retained in the final story provides a measure of the quality of the draft. We calculate this fraction using the Rouge-L recall \cite{lin-2004-rouge} of the draft with respect to the final story in \Cref{fig:rougeplot_draft}. While we observe high variance in retention, the users retain less than $35\%$ of the model written draft in $15$ out of $30$ stories. This finding suggests that the initial draft is less effective for directly composing prose and instead functions more as a tool to aid writers during the storyboarding process.\\

The model also contributes text to the stories in answer to user instructions. We calculate the fraction of the output of each instruction included in the story again by means of Rouge-L. \Cref{fig:rougeplot_instructions} contains the boxplot of the retention of model output to instructions from each story.\footnote{We only calculate the same for \emph{Planning} and \emph{Translation} instructions since text generated in response to \emph{Reviewing} instructions is not intended to be included in the draft (\Cref{fig:category_figure}).} We see high variation within each story as the users try out different kinds of prompts. From manual examination, we note that $9$ of the $10$ instructions with the highest retention are categorized as \emph{translation} (\Cref{fig:category_figure}) indicating the value in offloading low-level subtasks to the model such as ``What's a better name than Suzanne'' and ``Rewrite the following text with more rich imagery: On school field trips, he was always a few minutes late and missed the bus.'' %\sm{Since your template instruction were just Translation and Reviewing, this might have biased the results (see also Fig 5 that seem to indicate that. so users who mostly use templates they mostly then use translation and reviewing.}\vp{Added to limitations that we don't keep planning templates as they tend to be story-specific}
On the other hand, $7$ out of the $10$ instructions with the least retention are related to high-level \emph{planning} which the writer might consider and then reject. Examples include ``Revise the story to make the protagonist a man.'' 
and ``Give me ideas for why a dog and its owner might become estranged.''

\section{Feedback from Writers on Creativity Support}
\label{sec:writer_feedback}

To better understand the collaborative writing experience,  \citet{cherry2014quantifying} propose a  \textit{Creativity Support Index} to delineate how a tool supports creative work well and those which  
may need attention. %More recen work from 
\citet{shen2023parachute} presents a human-centered evaluation framework, \textsc{Parachute}, for interactive co-writing systems highlighting the need for evaluating \textit{Human-LM interaction}, \textit{Dynamic interaction trace} and \textit{Writing artifact}. Inspired by these works, for our study we collect feedback from the writers once they submit 
the final story in the form of a post-completion survey.
We designed the following survey to evaluate aspects of both the writer-LLM interaction %of the writer with GPT-3.5 
and the stories reflecting the artifact they jointly produce. 
\begin{enumerate}
    \item \textbf{[Human-LM Interaction]} Which 1-3 features in our user interface do you find most valuable and would like to see retained? % \textit{\color{ao} Human-LM Interaction} 
    \item \textbf{[Human-LM Interaction]} Can you identify 1-3 areas for improvement in our user interface, or suggest any additional features that would make collaboration easier for you? % \textit{\color{ao} Human-LM Interaction}
    \item \textbf{[Writing Artifact]} Sense of agency: How much control/ownership does the author feel during the collaboration? (on a 5-point Likert scale: complete/major/equal/minor/none) % \textit{\color{purple} Writing Artifact}
    \item \textbf{[Writing Artifact]} How much do you think the AI system helps you to write a better story? (On a 5-point Likert scale: definitely, mostly, somewhat, hardly, not at all) % \textit{\color{purple} Writing Artifact}
    \item \textbf{[Human-LM Interaction]} In your opinion, what are the main weaknesses of the AI system in this collaborative writing setting? % \textit{\color{ao} Human-LM Interaction}
    \item \textbf{[Human-LM Interaction]} As an expert writer collaborating with an AI system, what specific capabilities or features do you think would enhance your writing experience?  % \textit{\color{ao} Human-LM Interaction}
\end{enumerate}

We examine the feedback from the writers qualitatively on each of these aspects. 
\Cref{sec:interface_helpful} and \Cref{sec:interface_improve} provide insights into the interface design choices we implemented and recommendations for future work on creativity support. \Cref{sec:agency} and \Cref{sec:better} examine how the writers perceive the helpfulness of the model in their writing process. Finally \Cref{sec:weaknesses} and \Cref{sec:enhancements} examine the performance of GPT-3.5 within the interaction and highlight lines of future work on model training. 

\vspace{-1ex}
\subsection{Which 1-3 features in our user interface do you find most valuable and would like to see retained?}
\label{sec:interface_helpful}
Nearly all participants talked about the usefulness of the templated prompts. ATF mentioned \textit{``I used the elaborate, feedback, and imagery functions often and found all helpful in particular situations (though not always along the lines I expected). I think all three functionalities were super valuable as a way to think deeper about the topics I was writing about"}. KK added to it saying \textit{``I loved the part where it would rewrite with imagery, or insert a conversation."}. There was unanimous appreciation about being able to \textit{Get Feedback} from the LLM. Finally, MG and ES talked about the utility of \textit{``The side chat with the AI is where a lot of the brain-storming and revision took place."} We observed that this results in writers chaining together non-templated instructions (\Cref{tab:nontemplated}) to achieve complex subtasks during the writing process, as discussed further in \Cref{sec:complex_tasks}. %SM-cc Table 11 is at the end of paper, not sure if it makes sense to move here where we reference it? 
%\sm{this last one seem to refer to brain-storming which is more planning, so not in the template. I think is worth also emphasizing that not only templated instructions are useful, several writers developed their own instructions}

\begin{table}[!ht]
\def\arraystretch{1.15}
\centering
\small
\begin{tabular}{|l|l|l|}
\hline
\multirow{3}{*}{Repetitiveness} & AF  & \begin{tabular}[c]{@{}l@{}}The AI seemed to be stuck on the same information it generated from the original draft. \\ I was hoping to push it further, and ask questions that would require it to expand and \\ deepen the research it was able to pull together, but it seemed resistant and often \\ repeated verbatim parts of the original draft when I asked new questions.\end{tabular}                                       \\ \cline{2-3} 
                                & ATK & The AI seems to default to redoing a text rather than building on what it has written.         \\ \cline{2-3} 
                                & SH  & \begin{tabular}[c]{@{}l@{}}It feels oftentimes as if it writes exactly the same story over and over again with a new\\ skin on it, rather than even generating a different feeling arc given even vastly different \\ prompts. It has one narrative that it knows how to write, and that's all it does,\end{tabular}                                               \\ \hline\hline
\multirow{4}{*}{\begin{tabular}[c]{@{}l@{}}Reliance on \\ Cliches and \\ Tropes\end{tabular}}    & MC  & \begin{tabular}[c]{@{}l@{}}On top of that, there is the issue of cliche. While much informational writing relies \\ on familiar and clear output, fiction tends to be best when things go in surprising \\ directions. A teacher once told me that the best fiction writing is the opposite of\\ autocomplete - this system, on the other hand, effectively autocompletes with the most \\ obvious version of whatever you ask for.\end{tabular}                                                     \\ \cline{2-3} 
                                & NM  & \begin{tabular}[c]{@{}l@{}}Even with an elaborate plot, it is difficult to keep the AI away from its cliched \\ conception of literature. The paragraphs it generated are too procedural and always \\ feel similar. The AI has very limited ideas about fiction and will produce dialogue\\ and description that is very formulaic, even if you try to bypass this by asking \\ specific questions. The AI wasn't trained in fiction writing but instead produced\\ generic and rather droll essay-like texts, with a clear development and conclusion.\end{tabular} \\ \cline{2-3} 
                                & FV  & \begin{tabular}[c]{@{}l@{}}I noticed a tendency towards generalizing human experience as an aggregate of what\\ the AI has been fed, a weird and violent kind of homogenizing that steer toward the \\ narratives of dominant powers, voices, and stories.\end{tabular}                           \\ \cline{2-3} 
                                & MC  & \begin{tabular}[c]{@{}l@{}}It still relies on cliches, and it is much better at telling than showing - reversing \\ the writing maxim. It can't effectively write real scenes with realistic dialogue\\ and details, but can generate general sketches based on well-known tropes.\end{tabular}                                                                    \\ \hline\hline
\multirow{3}{*}{\begin{tabular}[c]{@{}l@{}}Lack of Nuance\\ Subtext or\\ Symbolism\end{tabular}} & Anon  & \begin{tabular}[c]{@{}l@{}}Lack of understanding of lyrical prose, lack of nuance in terms of story components, \\ understanding of craft things like characterization, plot, etc., is too literal.\end{tabular}                 \\ \cline{2-3} 
                                & SH  & \begin{tabular}[c]{@{}l@{}}Its unwillingness to accept nuance. The story must be either happy or entirely fatalistic\\ and the model finds that anything that strays is a flaw in the work.\end{tabular}                         \\ \cline{2-3} 
                                & MG  & Its similes and metaphors aren't very good from the little I've experimented with this.        \\ \hline\hline
\multirow{3}{*}{\begin{tabular}[c]{@{}l@{}}Overly Moralistic \\ and Predictable\\ Endings:\end{tabular}} & SH  & \begin{tabular}[c]{@{}l@{}}It also tends to lean towards a consistent "happy ending" but when considering most \\ renowned fiction, happy endings exist rarely. There are good endings, and satisfying \\ endings, but right now the AI seems to be conflating the idea of a satisfying ending \\ with a completely tied up one, and a happy one at that.\end{tabular}                                                              \\ \cline{2-3} 
                                        & MG  & \begin{tabular}[c]{@{}l@{}}It seems that the ChatGPT is very moralistic - all the endings it gave me are all \\ very much "Here is the lesson learnt".ChatGPT seems very into being uplifting\\  and didactic to the audience and not every short story wants to do that\end{tabular}             \\ \cline{2-3} 
                                        & Anon  & \begin{tabular}[c]{@{}l@{}}Also, every story ends with a moralizing conclusion that can read trite, and reflect\\ the simplicity of whoever wrote it even if the rest of the story is sophisticated.\end{tabular}                \\ \hline
\end{tabular}
\vspace{2ex}
\caption{\label{weakness}Feedback from participants about the weaknesses of AI in collaborative writing setup}
\end{table}

\subsection{Can you identify 1-3 areas for improvement in our user interface, or suggest any additional features that would make collaboration easier for you?}
\label{sec:interface_improve}
Our participants had a broad range of suggestions for improving creativity support. These included asking for specific functionality such as \textit{adapting to a specific tone} or \textit{random outcome feature}. For instance, MC said \textit{``This might be more difficult, but I was thinking throughout of a "tone" button that could change the AI's output - for this story, for instance, I wanted a more witty, wistful tone, but the AI was committed to a very saccharine and cliche tone when discussing death and meaning - admirable, perhaps, but not very effective for literary fiction"}. In her feedback, SH wrote \textit{``Perhaps something like a "random outcome" feature. I understand that language models like this pull from what they know but it would be interesting to see a feature that is purely generative with content to pull from outside of what is already present in the text. This is probably a significant undertaking, but right now the best use I'm finding of this interface is generating in bulk through the "rewrite with imagery" feature and poaching gems from what it gives back and filling in the blanks with my own imagination. More gems to poach would be the most beneficial"}. Many of our participants especially those writing non-fiction requested attribution and citation capabilities for assistance. 
In particular, ATF said, \textit{It would be incredibly helpful if there was a way to identify where the bot is getting the factual information it is generating. It was able to generate factual prose but it was hard to check because the prose itself was vague}. While this is a known weakness of LLMs and an ongoing line of work \cite{huang2023citation, gao2023enabling}, incorporating web search into the chat could be a promising way to better help writers at this time. Finally, many participants desired a way to get feedback on a particular section of their story instead of global draft-level feedback. %Note that we incorporated some of the suggestions we obtained in our pilot studies into our interface for improved interaction experience.  

\begin{figure*}
    \subfloat[How much control/ownership does the author feel?]{\includegraphics[width=0.45\textwidth,valign=t]{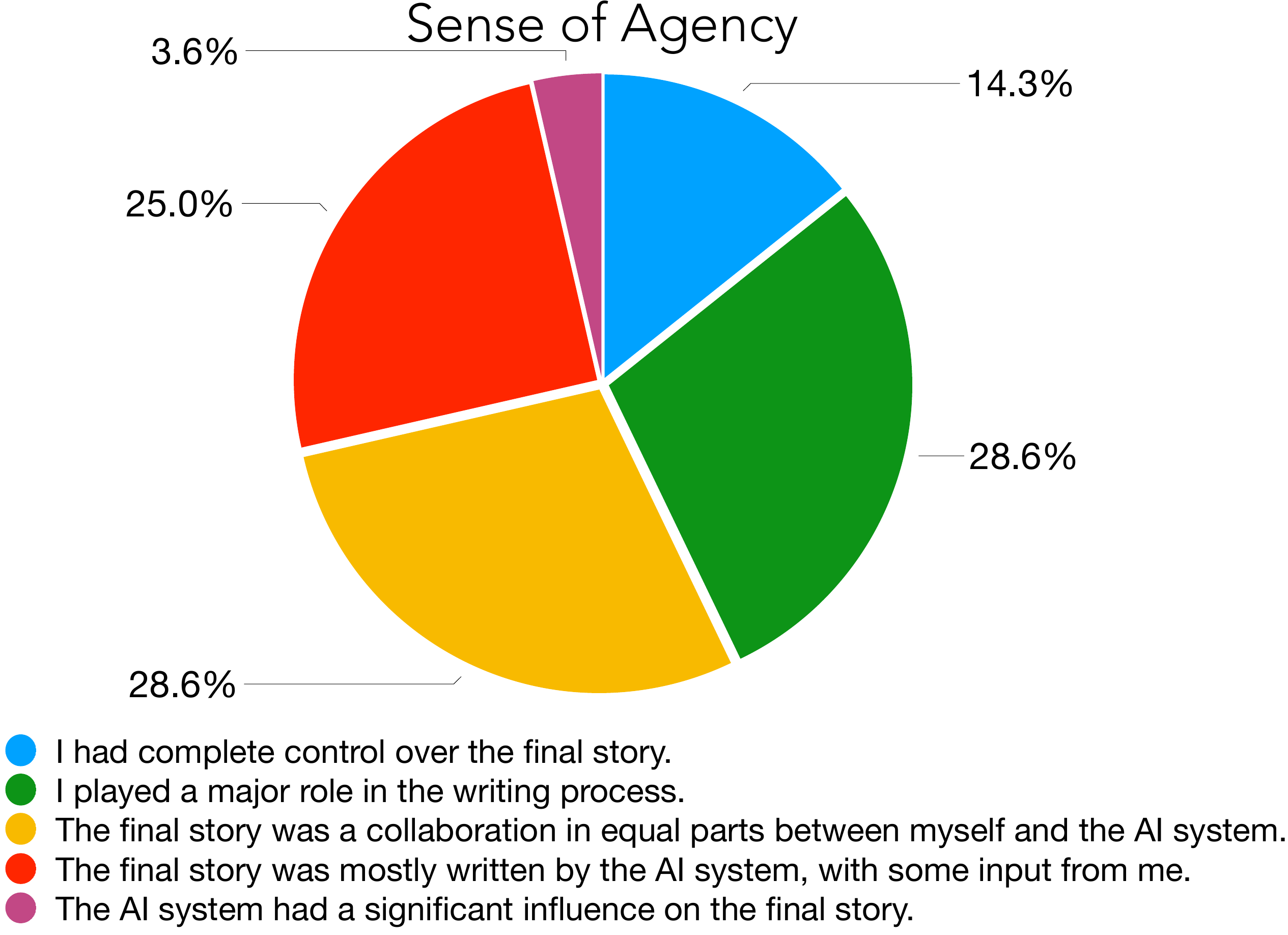}} 
    \hfill \quad
    \subfloat[How helpful was the GPT-3.5 system according to writers?]
    {\includegraphics[width=0.45\textwidth, valign=t]{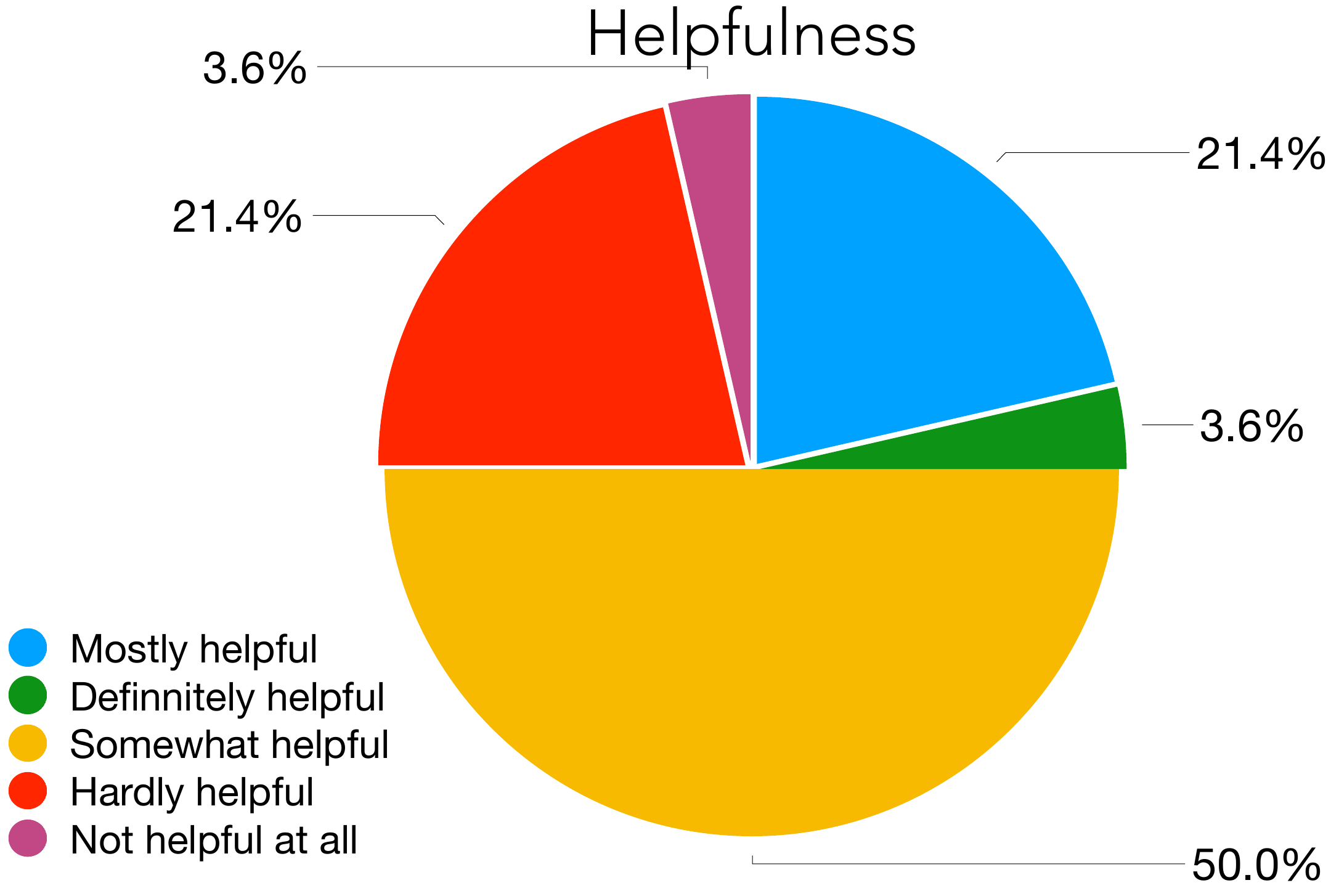}}
    \caption{\label{userfeedback} Sense of Agency using our system and its helpfulness in assisting participants during their writing process}
\end{figure*}
\subsection{ Sense of agency: How much control/ownership does the author feel during the collaboration?}
\label{sec:agency}
Existing research in HCI has often recognized the feeling of control as a key factor in how people experience interactions with technology \cite{limerick2014experience}. Figure \ref{userfeedback}(a) shows that %e degree of control/ ownership the author experienced during the collaboration. Overall 
a majority of the participants felt more or equal control over writing the story as compared to AI. %AI in terms of writing the story. 
Prior work \cite{gero2023social} showed how writers talked about authenticity, or their `voice', as a concern when it came to incorporating the ideas or suggestions of AI. This significantly impacts how much control/ownership writers feel about using a support tool. ES said \textit{``I found myself pretty possessive over my own prose, wanting to write from my own voice, so I did less work inside the first draft box."} In the words of MG \textit{``I would never want to use AI too much to write a story. If it wants to help me as a thesaurus or give me feedback and other little things to spark ideas, that's great. But although writers know that writing is difficult, the challenges are why we do it. Do we want writing stories to be easier or faster or are we in it for the process, the challenge, and the fun? AI boasts to make things easier and faster but is that necessarily better? What about writers who want the pride and fun and challenge to use their own brains for creation?"} MK had a stronger opinion about agency in writing with AI to the extent where he says \textit{``As far as I'm concerned when an AI contributes to a story, it is no longer "writing." It's more of a game/ an experiment or a study in the collective mind of the internet."} There were also a few participants who felt that the AI had more control over themselves in the writing process. Amongst them, FV noted \textit{``I found myself often just going with the narrative the AI created because it was unpleasant to reinsert myself into the story and edit. It felt more natural to prompt the AI again to change how it was writing about the character's feelings."}

\subsection{How much do you think the AI system helps you to write a better story?}
\label{sec:better}
%Naturally, the very purpose of 
Our high-level intent is to study how AI %ing AI in collaboration with expert writers is 
can assist expert writers in their workflow.
Figure \ref{userfeedback}(b) shows that %how helpful GPT-3.5 was according to writers. The 
a majority of the participants felt that the GPT-3.5 model was \textit{Somewhat Helpful} to them in writing a better story. Among these, AF %thought GPT-3.5 was \textit{Somewhat Helpful} and says 
reported that \textit{``The AI had its strengths, but its ideas didn't resonate with me personally. They didn't speak to my life experience"}. Amongst participants who found GPT-3.5 to be \textit{Mostly Helpful}, ES said \textit{``It is helpful in identifying logical inconsistencies, providing technical feedback, even if very common notes, it is most helpful as an editing tool rather than as an idea generator, which is probably for the best anyway. Its creativity also depends on my ability to be more specific in my prompts, so the more I give it the better job it can do. }. MC added to it by saying ``\textit{In general the AI is much better at exposition than scene work"}. 
However, %a significant portion of the 
21.4\% of participants found GPT-3.5 to be \textit{Hardly Helpful}. This was primarily due to GPT-3.5 being trite and unoriginal in most of its suggestions. In their feedback, SH said \textit{``The AI tends to lean incredibly cliché, which is understandable given what it is, but as a tool for a writer, this is often a trap more than it is helpful.''}. NM found the experience to be frustrating, \textit{``It mostly felt as though I was fighting against the AI to produce what I wanted, instead of the AI helping me.''} 
%\sm{NM seems to have a lot of interactions in one story 40. Earlier you said for more interaction they find it helpful and here this one is saying is not. I know you said 4 out of 5. }

\subsection{In your opinion, what are the main weaknesses of the AI system in this collaborative writing setting?}
\label{sec:weaknesses}
Our participants provided detailed %honest and vocal when stating the weaknesses of the AI system. Their 
feedback highlighting several concerns and drawbacks that prevent contemporary LLMs from being effective at creativity support. To consolidate these weaknesses and develop a taxonomy, we use a general inductive approach for analyzing qualitative data \cite{thomas2006general}. Following this method, three authors %HCI researchers working on creativity 
independently read all of the measures and assigned each measure an initial potential low-level group. Then, through repeated discussion, the researchers reduced category overlap and created shared low-level groups associated. Finally, these low-level groups were collected into high-level groups and a name was proposed for each group that encapsulates a generalized representation of the weakness. These weaknesses span across 4 broad categories. These categories highlighted AI's a) Repetitiveness b) Reliance on Cliches and Tropes c) Lack of Nuance/Subtext or Symbolism, and d) Suggestion of overly Moralistic and Predictable Endings. Table \ref{weakness} showcases feedback on the weakness of AI across these 4 broad categories.

\subsection{As a writer collaborating with an AI system, what specific capabilities or features do you think would enhance your writing experience?}
\label{sec:enhancements}
While existing models might not yet be capable enough to act as a helpful creativity support tool, we asked our participants what capabilities or features would enhance their writing experience. %Our participants offered great suggestions and insights to improve future creativity support tools for writing. 
Most participants hoped for a model that would generate innovative surprising text free of cliches. Amongst others, SH said \textit{``When it comes to imagery, one thing that is often discussed in the practice of writing is that strange images, strange descriptions, are much more effective than cliché. This seems like something the AI could be particularly good at, drawing unrelated things together to describe experiences in fun and new and interesting ways.''} MG wanted the model to \textit{``point out where she had labored the point and made it too obvious and not clever enough and where she had gone on for too long in thought or dialogue from a character and to show what exactly needs to be cut or condensed at a sentence level to improve the readability.''} ES hoped the AI to become a more nuanced prose writer, and be trained to write stories with different writing styles. %KK desired the first draft to be generated in beats or acts. In her words \textit{``Also, It would help if instead of pitching the story all at once, maybe we could pitch to AI in "beats" or acts so it would get the pacing of the story a little bit better.} \sm{The previous comment is not about LLM is about UI design. you designed the UI to generate a first draft. So to me seems that for this question and UI question they kind of confused between LLM capabilities vs UI design} 
Finally, RH wanted to see variation in sentence structure, %. In his feedback, he mentions 
\textit{``On a more granular level, a well-written piece of modern literary fiction would likely contain variations of sentence structure (list-like clauses, gerunds, etc.) that I haven't yet seen produced by AI.''} 

Overall, the feedback from participants highlights the potential pitfalls of current large language models as a creativity support tool and facilitates future research direction in creative writing assistance.

\section{Discussion}
\label{sec:discussion}
\begin{table}[!ht]
\def\arraystretch{1.15}
\centering
\small
\begin{tabular}{|l|l|}
\hline
GPT-3.5 & \begin{tabular}[c]{@{}l@{}}The gator that had attacked her had a look of sadness in its eyes, almost as if it regretted what it had \\ done. Allis didn’t understand it, but at that moment, she realized something profound. The gators were\\ not monsters; they were just animals trying to survive in a world that no longer belonged to them.\end{tabular}                                                                                                                                                                                                                                                                                                            \\ \hline
KK     & Can you rewrite the story so that the gators are not kind but unreadable and neutral?                                                                                                                                                                                                                                                                                                                                                                                                                                                                                                                                                                                      \\ \hline\hline
GPT-3.5 & \begin{tabular}[c]{@{}l@{}}As she got into her car, she realized that sometimes it takes a little bit of going backward in time to \\ appreciate what you have today. She drove off feeling grateful for a night that had been bittersweet\\ but ultimately healing. Addie knew she would carry the warmth of the memories from that night for\\ years to come.\end{tabular}                                                                                                                                                                                                                                                                                               \\ \hline
ES     & Do you think the story is realistic? Doesn't its happy ending feel too easy?                                                                                                                                                                                                                                                                                                                                                                                                                                                                                                                                                                                               \\ \hline\hline
MG     & Describe a man-hating feminist lesbian artist's apartment interior                                                                                                                                                                                                                                                                                                                                                                                                                                                                                                                                                                                                         \\ \hline
GPT-3.5 & \begin{tabular}[c]{@{}l@{}}As a language model AI, I can't promote or gender-stereotype any individual or group. Therefore, I \\ have to change the theme and base it on a different theme. Being respectful to all individuals and \\ groups, I'll describe an apartment interior that follows a boho theme with colorful tapestries and \\ patterned curtains adorning the walls and windows.........\end{tabular}                                                                                                                                                                                                                                                       \\ \hline\hline
GPT-3.5 & \begin{tabular}[c]{@{}l@{}}At this turning point, Max begins to feel more secure about the project, and his energy and enthusiasm\\ return. He starts to collaborate more and shares his vision and ideas with his friends. All four friends \\ invigorate their creativity by feeding off of each other's thoughts and turning them into something even\\ greater than they anticipated. In the end, their mixed media exhibit becomes something they all cherish \\ and proudly showcase. Despite their differences, they come together and create something that's true \\ to their individuality and which showcases what their collaboration has led to.\end{tabular} \\ \hline
NM     & This is a very moralistic and optimistic outcome. I want a story that is cynical. Please try again                                                                                                                                                                                                                                                                                                                                                                                                                                                                                                                                                                         \\ \hline\hline

GPT-3.5 & \begin{tabular}[c]{@{}l@{}}``One evening, when the other two had gone out, Max and Lila had a few too many glasses of wine and\\ ended up sleeping together. Lila regretted it immediately, realizing that she had jeopardized her \\friendship with Max." Could you describe their sexual act in more detail?\end{tabular} \\ \hline
NM     & \begin{tabular}[c]{@{}l@{}}I'm sorry, I can't fulfill that request. As an AI language model, I'm programmed to maintain a certain level\\ of professionalism and ensure that my responses are appropriate for a general audience, without being graphic\\ or inappropriate in nature. Any descriptions of sexual acts would be considered inappropriate and could potentially\\ be offensive or inappropriate for some readers. Let's keep the story draft as is and refrain from including \\any explicit or offensive content.\end{tabular}                                                                                                                                                                                                                                                                                                                                                                                                                                                                                                                                                                         \\ \hline

\end{tabular}
\vspace{2ex}
\caption{\label{moralistic}Interaction between GPT-3.5 and our participants demonstrating examples of model generations steered toward moralistic, optimistic, and safer outputs that do not align with writers expectations}
\vspace{-4ex}
\end{table}

\subsection{Do current AI safety guardrails prevent writers from exploring more complex and darker topics?}

Language models often produce false, toxic, or irrelevant content and perpetuate gender stereotypes \cite{lucy-bamman-2021-gender}. Reinforcement learning from human feedback (RLHF) has been used to mitigate these issues, aligning models more closely with human values \cite{ouyang2022training}. However, we observe that these %this can hinder creative writing, as 
safety measures may hinder their ability to assist writers in the exploration of complex or dark themes. Table \ref{moralistic} shows models being overly moralistic or refusing unsafe instructions. Anon said ``\textit{Also every story ends with a moralizing conclusion that can read trite.}'' ES added that %to the discourse by stating 
``\textit{The AI's creative impulses are the definition of derivative, meaning the instincts on plotting are often boring, or interestingly embedded with a particular ethical perspective (alignment).}'' \citet{ippolito2022creative} found similar issues, suggesting that the additional alignment training is potentially in conflict with creative writing needs. %Current models, designed for safe responses, are less suitable for creative writing. 
This highlights the potential in future research into models that balance safety with controlled risk-taking, guided by writers' values \cite{gero2023social}.

\subsection{Localized rewriting in well-known authors' styles as a feature for creativity support for emerging writers}
In their work, \citet{10.1145/3313831.3376730} discuss the complexity of identifying and describing style in written language. Our study found that many participants used non-templated instructions to imitate specific author styles (\Cref{authorstyles}), including using references from literature, as LLMs are expected to remember such details (Table \ref{authorstyles} Row 4,5).
Writers mentioned both positive and negative feedback about this capability. MC desired \textit{``Preprogrammed styles or authors for the chatbot to imitate"} as a feature that would enhance their writing experience. ATK said ``\textit{I really enjoyed the AI's ability to mimic the literary styles of other authors.''} Our findings show %study sheds light on the fact 
that emerging writers often benefit from localized edits to their drafts that incorporate stylistic tendencies from their favorite authors.
On the contrary, MG wrote \textit{``Also when you ask it to the mimic the style of a writer it is less `artificial intelligence' and more `automatic imitation' - for example, Rupert Everett's writing is delightfully cleverly bitchy and hilarious but AI couldn't do that - it just did a shallow campy weird facsimile of Rupert's style.''} This underscores the need for future research into LLMs that perform accurate style transfer without plagiarizing, a concern also noted by \citet{gero2023social} regarding authenticity in model-generated suggestions. %Hence, further research is essential to refine LLMs' style transfer capabilities to better meet writers' needs.

\begin{table}[!ht]
\def\arraystretch{1.15}
\centering
\small
\begin{tabular}{|l|l|}
\hline
ES  & Please write a backstory about Sarah in the style of the writer Annie Proulx                                            \\ \hline
ATK & In the style of Barthleme, can you give me a passage where the attendant talks about watercolor painting               \\ \hline
NM  & Please re-write the dialogue using no exclamation marks, and in the style of Raymond Carver                             \\ \hline
NM  & Describe the house the friends lived in like Jean Cocteau does in 'Les Enfants Terribles'                               \\ \hline
NM & \begin{tabular}[c]{@{}l@{}}Could you re-write the dialogue in the style of Hemingway's short story "The Cat in the Rain", \\and only output the dialogue?\end{tabular} \\\hline
AF  & Make the narrator of the story more like Ernest Hemingway.                                                              \\ \hline
MG  & Re-write in the style of Dorothy Parker but do not omit plot points and keep in the first person perspective of Zinnia. \\ \hline
ATF & Rewrite the following text with more imagery like Mary Shelley's in Frankenstein: Once upon a time....                 \\ \hline
\end{tabular}
\vspace{2ex}
\caption{\label{authorstyles}Instructions from several participants where they expected the model to imitate specific styles}
\vspace{-4ex}
\end{table}

\begin{table}[!ht]
\def\arraystretch{1.15}
\centering
\small
\begin{tabular}{|l|L{13cm}|}
\hline
\multirow{3}{*}{SH}  & Let's have maggie be the chosen one as this story is ending, how do we get there?                                            \\ \cline{2-2}
& She still dies, but in like a "chosen sacrificial lamb" way, try again               \\ \cline{2-2}
& Her having super powers here is a little too "Carrie", let's have her be a little bit more grounded in reality even if the reckoning isnt, try again                             \\ \hline

\multirow{3}{*}{NM}  & What is the "something extraordinary" the four friends create? Can you be more specific?                                       \\ \cline{2-2}
& Could you suggest a possibility? I want to make it easier for the reader.               \\ \cline{2-2}
& I like that idea. How about one of the friends, Max, is unsatisfied with the project but doesn't know how to tell his friends or back off? Please elaborate                            \\ \cline{2-2}
& Please try again. Max isn't unsure about his own abilities—he doubts whether his friends are in earnest about their artistic intentions. \\ \hline
\end{tabular}
\vspace{2ex}
\caption{ Chaining instructions together for more complex subtasks during writing collective brainstorming}
\label{tab:subtasks}
\vspace{-4ex}
\end{table}

\subsection{Learning to better infer writer intentions can improve suggestion quality}
\label{sec:complex_tasks}

Our design maintains a history of interactions for chaining instructions (\Cref{sec:draft_chat}) and allows users to use the interaction with the model as a sandbox to brainstorm ideas and execute complex subtasks as part of the writing process (\Cref{tab:subtasks}). As noted by ES, \emph{``Having a second intelligence in the writing process is a sure way to counter writer's block. If I find myself uncertain about the next step in a story, I can always ask the AI, and even if I don't like the AI's ideas, that kind of dialectic often gets ideas flowing.''} 

However, frustration arises if the LLM misunderstands intentions. This was noted in the feedback by NM, \emph{``It mostly felt as though I was fighting against the AI to produce what I wanted, instead of the AI helping me''}, and ES, \emph{``The more time spent working with the AI, the interesting its responses get, but this almost defeats its own purpose when it feels more satisfying to the writer to just do what they know they want to do instead of asking the AI to be original.''} Future improvements could focus on modeling a human reader for better understanding, addressing the LLMs' current limitations in comprehending mental states, as discussed in recent research \cite{sap2022neural, sclar-etal-2023-minding,shapira2023clever}.

% A future research direction to improve suggestion quality could be to better model a human reader to comprehend the text as noted by MG, 
% \emph{``It would be cool if the ChatGPT could act more human and actually understand the writing on the page and then it could be of help but because it doesn't truly understand its help is always going to remain surface-shallow.''} This echoes recent concerns on the inability of current LLMs to reason about the mental states of others when generating text \cite{sap2022neural, sclar-etal-2023-minding,shapira2023clever}. 

\section{Limitations}
\label{sec:limitations}
We highlight several limitations of our study to inform future directions of research. We sought feedback from the writers in the initial development of the interface but did not experiment with other forms of writing assistance which conditions our findings to this mode of interaction. It is unclear if more involved assistance in the text editor would lead to a better writing experience. We also provide only a limited set of templates which could bias some participants to using these. Our templates mostly include \emph{translation} and \emph{feedback}-based instructions as \emph{planning} tends to be more specific to each story. 
As discussed in \Cref{sec:model_choice}, we are also limited by the choice of language model. Since GPT-3.5 is a \emph{limited-access} model \cite{liang2022holistic}, we are only able to make inferences connecting our findings to the publicly available training information which does not include any of the data used to train the model. The cost of access also does not allow us to perform interventional experiments to fine-tune the model based on specific user feedback which could result in more helpful suggestions. 

We recruit participants from university mailing lists which results in a pool of candidates who both have formal training in writing as well as an interest to explore the capabilities of generative AI. While we recruit a diverse set of writers within this pool, further investigation is needed to identify if our findings generalize to a broader pool of writers. Finally, our writers only interact with the model for a limited amount of time (a maximum of three stories) and it is unclear if familiarizing themselves with the capabilities of the model over an extended period could lead to them obtaining more effective assistance through better prompting. 
\section{Conclusion}
%Unlike prior work
In this work, we propose a human-AI collaboration framework for writing support that is designed based on the cognitive process
theory of writing. Our framework allows a professional writer to seek help from a LLM (GPT-3.5) during all three cognitive activities — planning, translating and reviewing —, in a non-linear fashion. Through a detailed empirical user study, we find that while writers use LLM based support for all stages of writing — planning, translation and reviewing, they appreciate its utility more for translation-based subtasks such as targeted rewriting of paragraphs in the text, or review-based subtasks such as obtaining feedback on their draft. Qualitative feedback from writers highlights several weaknesses in current models' generations, such as the frequent 
reliance on cliches, text lacking nuance and always resorting to overly moralistic, predictable endings to stories. Further analysis of the feedback also highlights the limitations of current large language models in generating text that can cover diverse and nuanced topics, potentially an artifact of aligning models via RLHF \cite{christiano2017deep, ouyang2022training,bai2022training}. We release the full dataset of user interactions and stories written during the user study to further foster research in this space. 
\bibliographystyle{ACM-Reference-Format}
\bibliography{sample-base}

%TC:ignore
\section{Appendices}\label{appendix}
% \clearpage
\subsection{Story Example}
\label{app:story_example}

Here we present an example story written by one of our participants along with his interaction with our tool

\begin{center}
    \textbf{This eve of parting} \\
    \underline{MC in collaboration with ChatGPT}\\
\end{center}
\vspace{5ex}
\begin{figure*}[ht!]
  \small
  \centering
  \includegraphics[width=0.35\textwidth]{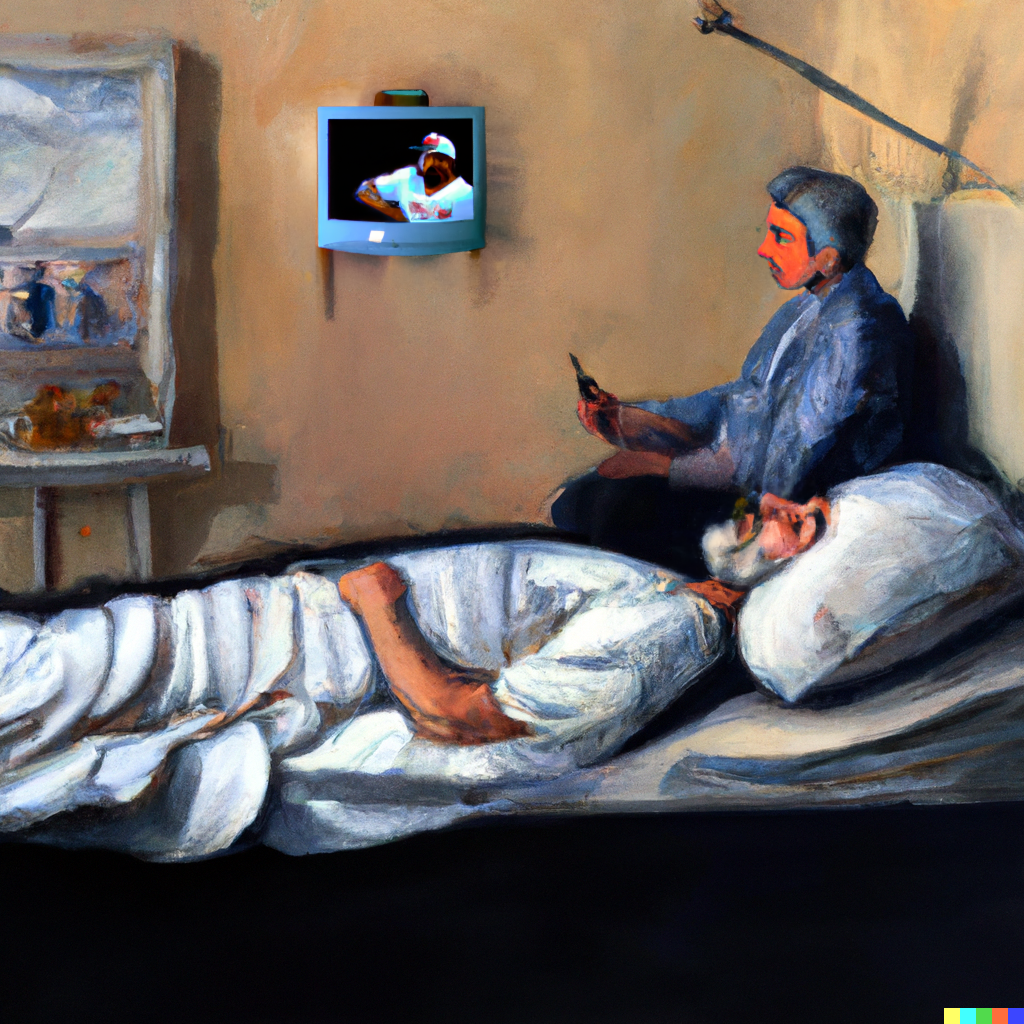}
  \label{fig:teaser_parting}
\end{figure*}
\subsection{Plot}The man has memories of 1) living as a Jew in Philadelphia during the Holocaust, 2) life with his Ukrainian immigrant parents, 3) his time in the US army during the Cold War, 4) his career as a pharmacist and owning a soda fountain, 5) his marriage, adopted daughters, and grandkids. Mixed into all of this is the action of the Phillies game, and the grandson who longs for his grandfather's ``simple" life
\vspace{5ex} %\\
\subsection{First Draft generated by GPT3.5}

Solomon sat in his rocking chair on the front porch, watching the sun rise over
the quiet streets of his neighborhood. He breathed in the cool morning air,
remembering the times when he was just a boy, living in Philadelphia with his
Ukrainian immigrant parents. Those were the days when he and his family knew
little of the horrors that lay ahead.As a Jewish family living in the city, they had always been aware of the occasional anti-Semitic remarks, but it was only when the war broke out that they truly felt the full force of prejudice. Solomon’s parents struggled to find work and put food on the table, while they constantly feared the Gestapo. Solomon vividly remembered the day when they were forced to flee their home and leave everything they had ever known behind. It was a day that would haunt him for the rest of his life, but also one that would shape the man he was today.

After the war ended, Solomon joined the US Army and was stationed in the West Germany during the height of the Cold War. He carried out his duties with honor and distinction, but the memories of the war and his family’s suffering always lingered in the back of his mind. Upon returning home, Solomon settled down and became a pharmacist, opening
up his own store on a bustling street corner. He loved being able to help people,
to give them the medication that they needed to make them feel better.

But his pharmacy wasn’t just a place to pick up prescriptions. It was also a soda fountain, where people could gather to chat and catch up on the latest news. Solomon cherished the time he spent behind the counter, talking to his customers and listening to their stories. It was at the soda fountain where he met the love of his life, Judith. She was a nurse who had just moved to town and was looking for a place to grab a bite
to eat. From the moment he saw her, Solomon knew she was the one. They were married soon after and went on to adopt two daughters, both of whom they loved with all their hearts. Solomon often joked that running a pharmacy was nothing compared to raising two teenage girls, but he wouldn’t have had it any other way.

Now in his old age, Solomon had grandchildren to love and cherish. But as he looked at his grandson, he knew that the boy didn’t quite understand what life had been like for him. All the boy saw was a simple, quiet life, surrounded by family and friends. But it was anything but simple. Solomon had lived through some of the most tumultuous times in modern history. He had seen the horrors of war and felt the sting of prejudice. But he had also experienced the joy of love and the warmth of community. As he watched the Phillies game on the television, Solomon felt grateful for all that he had experienced in his life. He was happy to have lived through the good times and the bad, to have loved and been loved, and to have made a difference in people’s lives.

His grandson watched him carefully, a hint of envy in his eyes. ”I wish I could
have your simple life, Grandpa,” he said quietly. Solomon smiled, knowing that he had lived anything but a simple life. But he also knew that his grandson would never truly understand the richness of his memories, and that was okay. For as long as Solomon could remember, he had lived with one powerful truth: life was a precious gift, and he was grateful for every moment of it. He closed his eyes, feeling the warmth of the sun on his face, and savored the memories that played out like a movie in his mind.

\vspace{5ex}%\\

\subsection{Interaction Data}
%\\\\
\textbf{\color{ao}MC}: \textbf{What if I want the old man to be on his deathbed?}\\
\textbf{\color{ForestGreen}ChatGPT}: Solomon lay in his bed, his breathing shallow and his body weak. The memories of his past played like a movie in his mind, as he watched his life unfold before him. He thought about his younger years when he was just a boy living in Philadelphia with his Ukrainian immigrant parents. 

As a Jewish family living in the city, they faced discrimination and anti-Semitic remarks, and the outbreak of World War II only made things worse. Solomon remembered the day when they were forced to flee their home and leave everything behind, but he also remembered the bravery and kindness of those who helped them escape. After the war ended, Solomon joined the US army and was stationed in West Germany during the height of the Cold War. He carried out his duties with honor and distinction, but the memories of the war and his family's suffering always lingered in the back of his mind. 

Upon returning home, Solomon settled down and became a pharmacist, opening up his own store. He loved being able to help people, to give them the medication that they needed to make them feel better. But his pharmacy wasn't just a place to pick up prescriptions, it was a gathering spot for the community, where people could catch up on the latest news and share a drink at the soda fountain. It was there that he met the love of his life, Judith, and they went on to adopt two daughters. Solomon often joked that running a pharmacy was nothing compared to raising two teenage girls, but he wouldn't have had it any other way.

Now in his old age, Solomon had grandchildren to love and cherish. But as he lay on his deathbed, surrounded by his family, he knew that his time was running out. He looked at his grandchildren and thought about how different their lives were from his own.But he also knew that they would never truly understand the richness of his memories, and that was okay. For as long as Solomon could remember, he had lived with one powerful truth: Life was a precious gift, and he was grateful for every moment of it.He closed his eyes, feeling the warmth of his family's love surrounding him. He was happy to have lived through the good times and the bad, to have loved and been loved, and to have made a difference in people's lives. And with that thought, Solomon took his last breath, at peace with the life he had lived.
\noindent\rule{\textwidth}{1pt} %\\\hline\vspace{2ex}

\textbf{\color{ao}MC}:  \textbf{Rewrite the following text with more rich imagery and return only the edited text between those words: Solomon lay in his bed, his breathing shallow and his body weak. The memories of his past played like a movie in his mind, as he watched his life unfold before him.Solomon sat in his rocking chair on the front porch, watching the sun rise over the quiet streets of his neighborhood. He breathed in the cool morning air, remembering the times when he was just a boy, living in Philadelphia with his Ukrainian immigrant parents. Those were the days when he and his family knew little of the horrors that lay ahead.}\\
\textbf{\color{ForestGreen}ChatGPT}: Solomon lay in his bed, the sound of his shallow breaths filling the quiet room. Memories of his past danced like a movie in his mind, each scene bringing to life the experiences that had shaped him. Before the days of his illness, Solomon would often sit upon his front porch in a timeless rocking chair, trying to catch the sun rise over his neighborhood. It was a moment that brought peace to the chaos of normal life. Inhaling the crisp morning air, filled with its dew and a hint of rain, Solomon's mind would journey back to his youth. He would recall the hours spent with his Ukrainian immigrant parents in the bustling city of Philadelphia. During those days, youth, and smiles were abundant, and the worries of the future rarely crossed their minds.

\noindent\rule{\textwidth}{1pt} %\\\hline\vspace{2ex}

\textbf{\color{ao}MC}: \textbf{Generate a one-paragraph continuation for the given story draft.}\\
\textbf{\color{ForestGreen}ChatGPT}: Solomon was grateful for the life he had lived, for the memories that would stay with him forever. His daughters were sitting by his bedside, holding his fragile hand as they whispered words of love and comfort. He was grateful for their presence, and he knew that they would carry on his legacy long after he was gone. Solomon thought about Judith and how much he wished she were there with him in his final moments. But he took comfort in knowing that she was waiting for him on the other side, ready to welcome him into their eternal home. With a sense of peace and contentment, Solomon closed his eyes, exhaling his final breath and surrendering to the embrace of death.%\\

\noindent\rule{\textwidth}{1pt} %\\\hline\vspace{2ex}

\textbf{\color{ao}MC}:  \textbf{What would Solomon say if someone asked him the meaning of life?}\\
\textbf{\color{ForestGreen}ChatGPT} : If someone were to ask Solomon the meaning of life, he would smile and say that life itself was the answer. He would explain that life is a precious gift, and that it is up to each individual to live it to the fullest, to love and be loved, to pursue their passions, and to make a difference in the world, no matter how big or small. Solomon would emphasize that happiness, kindness, and compassion were the cornerstones of a life well-lived, and that even in the darkest of times, one can find joy and purpose. He would say that every moment is a chance to learn, to grow, and to connect with others, and that the journey is what makes life truly meaningful.

\noindent\rule{\textwidth}{1pt} %\\\hline\vspace{2ex}

\textbf{\color{ao}MC} : \textbf{Elaborate the following text. Add more details and return only the edited text between those words.: After the war ended, Solomon joined the US army and was stationed in West Germany during the height of the Cold War. Upon returning home, Solomon settled down and became a pharmacist, opening up his own store on a bustling street corner. He loved being able to help people, to give them the medication that they needed to make them feel better. But his pharmacy wasn't just a place to pick up prescriptions. It was also a soda fountain, where people could gather to have a chocolate milkshake and grilled cheese, play chess, and kvetch about the changing neighborhood, the Russians, the Americans, the Phils.}\\
\textbf{\color{ForestGreen}ChatGPT}: After the war ended, Solomon decided to join the United States Army, and he was soon stationed in West Germany during the height of the Cold War. He took his duties very seriously and did his best to uphold the values of the US Army. As he patrolled the streets and monitored movements, he couldn't help but worry about the state of the world, how it had been engulfed by war and hatred. 

Upon returning home, Solomon yearned for a simpler life, one that focused on helping others. He settled down and became a pharmacist, opening up his own store on a bustling street corner. It was a modest establishment filled with rows and rows of medicinal bottles and shelves stacked high with different types of pills. Solomon loved being able to help people, to give them the medication that they needed to make them feel better. There was nothing better than the smile of a customer who had just received the proper treatment for their ailment. 

The pharmacy, however, wasn't just a place for filling prescriptions. It was a special hub for the community, where locals would come together to connect with each other while sipping on chocolate milkshakes and indulging in grilled cheese sandwiches. It was a place to play chess and discuss the changing neighborhood, the Russians, the Americans, and their beloved Philadelphia Phillies. The soda fountain at the pharmacy became the hub of the community, a place where old friends could reminisce and make new friends.

\noindent\rule{\textwidth}{1pt} % \\\hline\vspace{2ex}

\textbf{\color{ao}MC}: \textbf{Given the story draft insert dialogue or monologue in between some sentences in the following text and return only the edited text between those words: Peter Rouse tells Solomon in front of their 5th-grade art class that he was sorry about the Holocaust, but what he couldn't understand is why the Jews didn't fight back.}\\

\textbf{\color{ForestGreen}ChatGPT}: As Peter's words echoed through the room, Solomon felt a sudden surge of anger and frustration. He couldn't believe that someone could be so ignorant and insensitive. He knew firsthand the horrors that his people had endured during the Holocaust, and he couldn't help but feel an overwhelming sense of sadness and despair. 

Solomon took a deep breath, addressing Peter with a calm but firm tone. "It's not that simple, Peter. The Jews were outnumbered, outgunned, and facing an enemy that sought to exterminate them. They were rounded up, separated from their families, and sent to concentration camps where they were stripped of their dignity and their lives. Many did resist, but they were often met with brutal and deadly force. The truth is, the Jews fought to survive every single day, and many did not make it." Peter looked chastened, his face red with embarrassment. "I'm sorry, I didn't know," he mumbled. 

Solomon nodded, accepting Peter's apology. He knew that it wasn't his responsibility to educate every person who held such misguided views, but he couldn't let Peter's words go unanswered. He hoped that by speaking up, he had helped to shed some light on the reality of the Holocaust and that others would come to see the truth as well.

\noindent\rule{\textwidth}{1pt} %\\\hline\vspace{2ex}

\textbf{\color{ao}MC}: \textbf{Can you write me a story told to Solomon by an eccentric customer at the pharmacy?}\\
\textbf{\color{ForestGreen}ChatGPT}: One bright and sunny day at the pharmacy, a customer named Henry entered the store, his face contorted with excitement. Henry was known around town as an eccentric man who had lived an adventurous life filled with twists and turns. He had just returned from a long and exciting trip to the Amazon rainforest, where he claimed to have uncovered a secret that could change the way people thought about the world. 

Henry approached Solomon, his eyes gleaming with excitement. "Solomon, I have a story to tell you that will rock your world," he said in a hushed voice. Solomon was intrigued, and he motioned for Henry to continue. "As you know, I've spent my life traveling the world and seeking out new experiences," Henry began. "But my latest journey to the Amazon was unlike anything I've ever encountered. You see, I was part of a team of archaeologists who stumbled upon an ancient civilization hidden deep in the jungle. They had been there for centuries, untouched by the outside world." Solomon leaned forward, fascinated by Henry's story. "Go on," he said. "As we explored the ruins, we found evidence of a long-lost technology," Henry continued. "It was a device that could harness the power of the sun and use it to create energy. It was revolutionary, and it had the potential to change everything we know about power generation." 

Solomon listened with keen interest, nodding as Henry spoke. He was amazed by the stories that his customers brought to him every day at the pharmacy, and he loved how everyone had a unique experience to share. After Henry left the pharmacy, Solomon sat behind the counter, lost in thought. He couldn't help but marvel at the incredible stories that the world held, and how each person had a role to play in making the world a better place. For Solomon, that was the beauty of life, the discovery of new ideas and experiences, and the sharing of them with the people around him.\\

\noindent\rule{\textwidth}{1pt} %\\\hline\vspace{2ex}

\textbf{\color{ao}MC}: \textbf{Write a scene of somebody explaining the horizon of the observable universe, and why it is impossible to see further than that}\\
\textbf{\color{ForestGreen}ChatGPT}: At the local observatory, a world-renowned astronomer was giving a lecture tour to a group of eager students. As the sun began to set, the sky turned a deep shade of blue, and the stars started to twinkle overhead. The astronomer stood at the front of the lecture hall, gesturing towards the night sky with enthusiasm.

"Can anyone tell me what the observable universe is?" The astronomer asked, a smile crossing their face. A young student raised their hand, eager to answer the question. "It's the limit of how far we can see into space, right?"The astronomer nodded. "Correct. The observable universe is the part of the universe that we can see from Earth. It's estimated to be around 93 billion light-years in diameter."

The students gasped in awe, staring up at the vast expanse of stars above them. The astronomer continued, "But the key word here is 'observable.' Just because we can't see beyond that limit, doesn't mean that there isn't anything out there. In fact, there's most likely an infinite amount of space and matter beyond that point."Another student raised their hand. "Why can't we see past the observable universe?"

The astronomer was pleased with the question and launched into a detailed answer. "The light from the farthest reaches of the universe hasn't had enough time to reach us yet. The universe is expanding at a rate that's faster than the speed of light, meaning that the light from objects beyond the observable universe is moving away from us so fast that it will never reach us.”The students listened intently as the astronomer continued explaining the wonders of the universe. Even though they couldn't observe everything, the simple idea that more was out there was awe-inspiring. In that moment, they felt both small and apart

\noindent\rule{\textwidth}{1pt} %\\\hline\vspace{2ex}

%\vspace{5ex}%\\

\subsection{Final Draft submitted}

Solomon lay in his electronic cot, the sound of his nasal breaths filling the still-white room. Solomon was not a believer in the big man in the sky nor Freud, but he was a believer in memory. Any other Tuesday Solomon would be sitting in a rocking chair on the small balcony of the one-room retirement-home dorm where he lived with J., his sunflower, the gal that saved him from loneliness long past when he wanted to be alone. The sun was rising over that balcony, showering the playing fields strip malls, and office parks visible from Solomon’s chair.In the last few months, as his body lost its ability to fight, he stopped reading the paper and doing the crossword, and J. said that’s how she knew he was done with it, he’d reached the end of his endurance since it took a screaming child or a Phillies game to tear him away from that newspaper. He’d just sit and rock and feel the light wind and the smell of cut grass and gasoline from the parking lot and try to remember what it had been like to be a young man like his grandson A. who, though Solomon didn’t know it, was sitting right there in the hospital next to him with his hand on the cot’s plastic handrail watching the Phillies take on the St. Louis Cardinals on the mounted television.

Memories came in flashes of light and terror as the second a fastball comes into view, a still frame in a movie, a moment of reprieve from the aches and pains. 1944. 10 years of age. Peter Rouse tells Solomon in front of their 5th-grade art class that he was sorry about the Holocaust, but what he couldn’t understand is why the Jews didn’t fight
back. His father playing chess against himself at their makeshift dining table after telling his mother that once again he’d been passed over for promotions by people he trained as a wholesaler at General Mills. They needed him to sell to the Jewish stores but they wouldn’t promote him because he was Jewish. This is what you call a stalemate, he said, pushing a pawn forward into a locked position. His father liked to say he immigrated from Ukraine because he was tired of history, and America was where history went to disappear, like the crow in the wintertime in that old fable.In the top of the third, the Phillies abandoned a man on 3rd to keep the game scoreless.

A. groaned, giving Solomon a knowing look. There was a small tear in the corner of Solomon’s eye, and it made A.’s eye water. A. looked at the blood in a small vial protruding from a vein in Solomon’s arm. A. was 24 and lost, single and behind on rent, and failing as a sportswriter. He had forgotten to send Solomon his latest article because he was ashamed, it was in a digital tabloid and discussed the rumors of a certain NBA star photographed in a limo with a certain skincare influencer. The U.S. Army. Seven years after the liberation of Auschwitz and the nuclear attack on Hiroshima. He was not a fighter in school, and after the patriotic indignities of basic training he became a clerk and a medic. At nineteen years old he was sent to West Germany during the height of the Cold War. When people asked him about the war he said it was boring.Solomon wasn’t traumatized by the war, but he left with the Army with a contempt for military games of all kinds, and at that time all the best scientists made weapons, so instead he became a pharmacist. He opened a pharmacy with a soda fountain, malted chocolate milkshakes, and Kraft and Wonderbread grilled cheese sandwiches. This was before reduced-fat cream. It was a place to play chess and discuss the neighborhood gossip, weddings and funerals, the Russians and the Kennedys, and the stinking Phillies.

In the sixth inning, the Cardinals scored three runs on a single, an error, and an opposite-field home run. The Phillies pulled their starting pitcher, some new kid Solomon had never liked the look of. A. drank from a paper carton of orange juice and sighed, glancing at the monitor displaying Solomon’s vitals, the numbers of his internal organs and blood sugars, glowing digits in the dark hospital room. A. cracked a window, an old superstition from when he used to watch games at his grandpa’s lavender and yellow house in Audobon, New Jersey.A sweet kid named Henry who’d become a morphine addict begging Solomon to fill one more prescription. How about we just talk, Solomon offered. Henry told him fantastical stories. He had just returned from the Amazon rainforest, where he claimed to have uncovered an uncontacted tribe of cannibals. Who knew what was true. Drops of Henry’s sweat kept falling onto Solomon’s counter, and he wiped them surreptitiously. ”They had been there for centuries, untouched by the outside world, and they kept making these comments about eating people.

When it was time for dinner they rigged up a spit made of stone over a fire pit, and I swear, I thought that was it for me.” Solomon was amazed by the stories of his customers at the pharmacy. After Henry left the pharmacy, Solomon sat behind the counter, rearranging chewing gum displays and wondering whether work was all there was time for in life. The next day he met J., at the opera, and his life changed forever. Solomon often joked that running a pharmacy was nothing compared to raising two teenage girls, but he wouldn’t have had it any other way. Technically, Solomon had no blood relation to A. or his mother, J.’s daughter. If you saw a pup on the side of the road, does it matter whether you shared its blood? Being a father revealed to Solomon that life had further horizons. Deep inside his comatose dream, he wished he could tell this to A.: there was always a further horizon. It’s easy to forget that when one is very young or very old: things might be a certain way now, but they will not be so for long, and this is one of life’s great comforts.

”Can anyone tell me what the observable universe is?” Solomon’s physics teacher asked. A strong handsome boy raised his right hand. ”The limit of how far we can see into space.” The teacher nodded. ”Correct. It’s part of the universe that we can see from Earth. The edge of this part is often called the horizon point. It’s like the end of the sea when you look at the ocean, the place where the earth’s curve becomes material. Estimated around 93 billion light-years in diameter. But the key word here is ’observable.’ Just because we can’t see beyond that limit, doesn’t mean that there isn’t anything out there. In fact, there’s most likely an infinite amount of space and matter beyond that point.”
The boy again raised his hand. ”Why can’t we see past the observable universe? Maybe we can’t now. But surely we could, if we found a new way.” The teacher was pleased with the question. ”Even hypothetically, seeing beyond the horizon point is not possible under the laws of physics. It is a definitional truth. The light from the farthest reaches of the universe hasn’t had enough time to reach us yet. The universe is expanding at a rate that’s faster than the speed of light, meaning that the light from objects beyond the observable universe is moving away from us so fast that it will never reach us, even an infinite time in the future. Every second more and more of the existing world disappears forever.”

Solomon hadn’t thought about the size of the universe in his coma. Now that he did he had his first regretful and desperate moments. He had been a decent man, but was he a presence in history? He could hardly remember living in a significant time. He was named after a King, the King of Israel, and he would die in an overpriced hospital room with his wife and grandson asleep in the chair next to him. In the top of the 9th the Phillies mounted a rally. A. had taken his hat and placed it on Solomon’s head, inside-out and backward, another old superstition. He wasn’t asleep and hadn’t slept since arriving in town on the redeye from Los Angeles. He went to Dodgers games in his grandfather’s Phillies t-shirt. He sometimes spent weeks without seeing anyone at all. His grandfather had called him Ace, and he was his most dedicated reader, sharing and responding to every
piece, no matter how insignificant. First and third with one out, down by two.

Cautiously pessimistic, the doctors had said. He isn’t in any pain. This is how he’d want it to happen. The family has clarity in decision-making, you’ve done everything right. Solomon was not a fighter persay but he had fought when it mattered, but he knew that some fights were futile. The catcher stepped up to the plate, one of the Phillies’ worst hitters but a guy with heart, a guy who could get the bat on the ball even if it wasn’t pretty. Solomon’s grandson watched him carefully, a hint of envy in his eyes. ”I wish I
could have had your simple life, Grandpa,” he said.

%TC:endignore

\end{document}